\documentclass{article}

\usepackage{arxiv}

\usepackage[utf8]{inputenc} 
\usepackage[T1]{fontenc}    
\usepackage{lmodern}        
\usepackage{hyperref}       
\usepackage{url}            
\usepackage{booktabs}       
\usepackage{amsfonts}       
\usepackage{nicefrac}       
\usepackage{microtype}      
\usepackage{graphicx}

\title{Double Standards: The Implications of ``Near'' Certainty Drone
Strikes in Pakistan}

\author{
    Shyam Raman
   \\
    Brooks School of Public Policy \\
    Cornell University \\
  Ithaca, NY \\
  \texttt{\href{mailto:sr2297@cornell.edu}{\nolinkurl{sr2297@cornell.edu}}} \\
   \And
    Paul Lushenko
   \\
    Brooks School of Public Policy \\
    Cornell University \\
  Ithaca, NY \\
  \texttt{\href{mailto:pal243@cornell.edu}{\nolinkurl{pal243@cornell.edu}}} \\
   \And
    Sarah Kreps
   \\
    Brooks School of Public Policy \\
    Cornell University \\
  Ithaca, NY \\
  \texttt{\href{mailto:sarah.kreps@cornell.edu}{\nolinkurl{sarah.kreps@cornell.edu}}} \\
  }

\newlength{\csllabelwidth}
\newlength{\cslhangindent}
  {}%
  {\par}
\newenvironment{CSLReferences}[2] 
 {
  \setlength{\parindent}{0pt}
  \ifodd #1 \everypar{\setlength{\hangindent}{\cslhangindent}}\ignorespaces\fi
  \ifnum #2 > 0
  \setlength{\parskip}{#2\baselineskip}
  \fi
 }%
 {}
\usepackage{calc} 

\usepackage{animate} \usepackage{booktabs} \usepackage[para,online,flushleft]{threeparttable} \usepackage{placeins} \usepackage{siunitx} \usepackage{xcolor} \usepackage{setspace} \newcolumntype{d}{S[input-symbols = ()]}

\begin{document}
\maketitle

\def\tightlist{}

\begin{abstract}
\singlespacing In 2013, U.S. President Barack Obama announced a policy
to minimize civilian casualties following drone strikes in undeclared
theaters of war. The policy calibrated Obama's approval of strikes
against the ``near'' certainty of no civilian casualties. Scholars do
not empirically study the merits of Obama's policy. Rather, they rely on
descriptive trends for civilian casualties in Pakistan to justify
competing claims for the policy's impact. We provide a novel estimate
for the impact of Obama's policy for civilian casualties in Pakistan
following U.S. drone strikes. We employ a regression discontinuity
design to estimate the effect of Obama's policy for civilian casualties,
strike precision, and adverted civilian casualties. We find a
discontinuity in civilian casualties approximately two years before
Obama's policy announcement, corroborating our primary research
including interviews with senior officials responsible for implementing
the near certainty standard. After confirming the sharp cutoff, we
estimate the policy resulted in a reduction of 12 civilian deaths per
month or 2 casualties per strike. The policy also enhanced the precision
of U.S. drone strikes to the point that they only killed the intended
target(s). Finally, we use a Monte Carlo simulation to estimate that the
policy adverted 320 civilian casualties. We then conduct a Value of
Statistical Life calculation, the first-of-its-kind in drone warfare
scholarship, to show that the adverted civilian casualties represent a
gain of 80 to 260 million U.S. dollars. In addition to conditioning
social and political outcomes, then, the near certainty standard also
imposed economic implications that are much less studied.
\end{abstract}


\break

\hypertarget{introduction}{%
\section{Introduction}\label{introduction}}

In a May 2013 speech at the National Defense University in Washington,
D.C., U.S. President Barack Obama announced the implementation of a
policy to exercise greater oversight of strikes conducted by Unmanned
Aerial Vehicles (UAV), or drones, against suspected
terrorists\footnote{{``Remarks by the President at the National Defense
  University''} (2013).}. The policy -- formally termed Presidential
Planning Guidance (PPG) -- responded to increasingly vocal concerns and
criticism from international and non-governmental organizations about
civilian casualties that resulted from the United States'
counterterrorism drone strike program. The PPG intended to minimize
collateral damage by imposing a near certainty standard for no civilian
casualties during U.S. drone strikes in undeclared theaters of war. To
what extent did the Obama administration's policy cause a reduction in
civilian casualties following U.S. drone strikes?

Where analysts have engaged this question, especially in terms of U.S.
drone strikes in Pakistan, their explanations are inconclusive. Scholars
rely on the same data compiled by one of several accountability
organizations to substantiate two opposing claims. ``By averaging the
high and low casualty estimates of militant and non-militant deaths
published in a wide range of reliable media outlets,'' Bergen and Sims
estimate ``that the civilian death rate in U.S. drone strikes in
Pakistan has declined dramatically'' (Sheehan, Marquardt, and Collins
2021). The proximate cause for Lindsay (2020) is that ``the Obama
administration piled on many layers of oversight, and error rates
decreased.'' Turse (2021), a freelance journalist, uses a similar
descriptive strategy to argue that ``mountains of evidence demonstrate
consistent failures.'' One prominent international legal expert adds
that the Obama administration's policy ``to impose stricter limitations
on drone strikes seems to have had an effect only at the margins''
(Jaffer 2016). These countervailing perspectives point to a broader gap
in the literature for the evolution of U.S. drone warfare following the
terrorist attacks of 9/11. Together, they suggest the need for an
empirical study to determine the causal significance of the Obama
administration's reform to help minimize civilian deaths following U.S.
drone strikes in undeclared theaters of operations.

The primary goal of this study is to determine the extent to which the
Obama administration's near certainty standard causally impacted the
rate of civilian casualties in undeclared theaters of war. Providing a
causal estimate for the implications of the Obama administration's near
certainty standard is important for a number of reasons. First, while
descriptive inference is useful to identify unobserved trends, it is
subject to endogeneity and cannot provide a causal estimate for the
impact of the near certainty standard (King, Keohane, and Verba 1994).
Evaluating the effects of policy on strike-level outcomes, then, becomes
challenging using methods that primarily rely on descriptive analysis
(Khan 2021; G. Martin and Steuter 2017; Sanger 2017; Shah 2018). Second,
we cannot determine the degree to which mistakes -- such as the Biden
administration's botched strike in August 2021 that killed ten Afghan
citizens rather than a suspected terrorist -- may be prevented by also
applying the near certainty standard in declared theaters of war (Aikins
et al. 2021). Therefore, our analysis contributes to ongoing policy
debates about how the U.S. should conduct strikes against terrorists
amid its withdrawal from Afghanistan and other conflict zones.

We provide a causal estimate for the implications of the near certainty
standard using data from the Bureau of Investigative Journalism (BIJ).
The BIJ dataset captures the universe of U.S. drone strikes in Pakistan
from 2002 to 2019 based on news reports, official statements, and
government press releases. According to senior U.S. policy officials we
interviewed for this paper, the BIJ is the most authoritative dataset
and provides information on the death toll and injuries resulting from
U.S. drone strikes for both adults and children\footnote{See the
  appendix for information on our elite interviews.}. We leverage the
BIJ data in a regression discontinuity (RD) design and exploit variation
at the Obama administration's implementation of the near certainty
standard to provide our estimate. Importantly, the implementation period
or cutoff -- informed by our multiple interviews with Obama-era
officials responsible for exercising the policy -- lies approximately
two years before the Obama administration's public announcement.

We find that the Obama administration's shift to the near certainty
standard 18 to 30 months prior to the official policy announcement
dramatically reduced civilian casualties in Pakistan. Consequently, the
policy markedly increased the precision of U.S. drone strikes, which
scholars often impugn for arbitrarily imposing risk based on the
``randomness'' or ``vagueness'' of targets (Aaronson et al. 2014; G.
Martin and Steuter 2017). Under the Obama administration, strikes
increasingly removed only the intended target(s). This finding suggests
that the promise of drone warfare -- reducing risk to a country's own
forces concomitant to better protecting civilians -- is possible but
requires important political and military trade-offs that officials may
be unwilling to make because they often impose greater risks on friendly
forces. Leveraging the causal interpretation of our RD estimate and the
``as-if'' randomness of civilian casualties, we also simulate a
projection of civilian casualties absent the policy. We estimate 320
averted civilian deaths attributable to the Obama administration's near
certainty standard. These represent a Value of Statistical Life (VSL)
gain of 80 to 260 million U.S. dollars (USD). This result suggests that
in addition to conditioning social and political outcomes, the certainty
standard governing U.S. drone strikes also imposes key economic
implications that are much less understood.

The remainder of this article unfolds in four parts. In Section 2, we
position the Obama administration's policy of near certainty within the
broader debate for U.S. officials' use of drone strikes since 9/11. In
Section 3, we discuss our empirical approach and regression
discontinuity in time (RDiT) specification. In Section 4, we discuss our
results and several robustness checks conducted following guidance by
Hausman and Rapson (2018). In Section 5, we discuss our findings and
their limitations, and conclude.

\hypertarget{background}{%
\section{Background}\label{background}}

\hypertarget{drones-and-u.s.-military-strategy}{%
\subsection{Drones and U.S. Military
Strategy}\label{drones-and-u.s.-military-strategy}}

In 2002, U.S. President George W. Bush authorized the first-known use of
an armed drone -- a General Atomics MQ-1 Predator -- to kill an al-Qaeda
leader in Yemen (Kreps 2016). Following the Bush administration's steady
increase in drone strikes, the Obama administration institutionalized
the weapon -- a practice that Donald Trump continued. Obama himself
viewed drones as a ``cure-all for terrorism''\footnote{{``Remarks by the
  President at the National Defense University''} (2013).}. On the heels
of the United States' withdrawal from Afghanistan, U.S. President Joseph
Biden adopted an ``over-the-horizon'' counterterrorism strategy that
suggests drone warfare will maintain its foothold in American military
strategy (Kreps and Lushenko 2021). The United States has used armed
drones in both active and non-active areas of conflict. Whereas the use
of force in active conflicts such as Afghanistan, Iraq, and Libya was
sanctioned by domestic or international organizations, the use of force
in places such as Pakistan, Somalia, and Yemen are a different matter.
These areas are characterized by operations that are neither sanctioned
by international organizations nor explicitly endorsed by domestic
authorization. To justify its use of drone strikes here, the U.S. has
cited several documents. These include the United Nations (UN) Article
51 right to self-defense, Article II statutory powers of the President's
Commander in Chief authority, or the 2001 Authorization for the Use of
Military Force that allowed American presidents to attack al-Qaeda and
its affiliates (Swan 2019).

The trajectory of U.S. counterterrorism drone strikes suggests two
dichotomous patterns of operations between declared and undeclared
theaters of war. The differences in these patterns are necessary to
properly account for the origins of the near certainty standard as well
as adjudicate its effectiveness over time. Drone strikes outside the
context of armed conflict have drawn the most criticism internationally
because they are conducted largely unilaterally and raise questions
about violations of sovereignty in ways that drone strikes within the
context of authorized armed conflict do not. Drone strikes also,
invariant of the theater of operations, threaten a ``spectrum of
impunity'' where civilians are exposed to greater dangers while
technologically superior combatants face far fewer battlefield risks
(Riza 2013). For some philosophers, drone strikes have now made
``\textit{combatant} immunity'' the guiding moral principle of war
(Primoratz 2007). Differences in the use and constraint of U.S. drone
strikes help explain why a ``human rights community'' consisting of
intergovernmental organizations (e.g., Office of the UN High
Commissioner for Human Rights), non-governmental organizations (e.g.,
Amnesty International), and academic institutions (e.g., Stanford
University's International Human Rights and Conflict Resolution Center)
systematically tracked U.S. drone strikes in Pakistan, Somalia, and
Yemen, while it was largely silent on Iraq and Afghanistan (Lushenko,
Bose, and Maley 2022). The collection of these data by watchdog
community organizations made apparent the large civilian toll which was
resulting -- in real time -- from the U.S. drone program.

Between 2002 and 2020, data from the BIJ estimates 10,000 to 17,000
people were killed by U.S. drone strikes in Afghanistan, Pakistan,
Somalia, and Yemen. Of these, between 800 and 1,750 were reported as
civilian casualties. The highest percentage of these non-combatant
deaths were in Pakistan. Roughly 14 to 17 percent of the deaths
resulting from U.S. drone strikes in Pakistan -- between 335 and 636
lives -- were reported as civilian casualties. The data provided by the
BIJ also shows 286 to 425 children killed by U.S. drone strikes in
Pakistan, which constituted 21 to 33 percent of the overall civilian
death toll.

The human toll of U.S. drone strikes is not limited to non-combatant
fatalities. Drone strikes impose long-term psychological, environmental,
and economic impacts on people living in targeted areas. These costs are
usually hidden and therefore much less acknowledged and studied by
researchers. U.S. drone strikes in communities near the
Afghanistan-Pakistan border are reflected in `` {[}\ldots{]} populations
experiencing immense stresses, reflecting fear of strikes, and
disruption to already marginal subsistence livelihoods based on culture
and limited trade {[}\ldots{]}'' (Page and Williams 2021). One
interviewee also explained ``we try to avoid going to each other's
homes.'' The interviewee continued by stating ``we try to avoid making
relations with newcomers or strangers, because it's hard to trust
everyone. We cannot recognise the person whether he is a terrorist or a
good person or if he is a target of drones or not'' (Page and Williams
2021).

In the early years of U.S. counterterrorism policy based on drone
strikes, the ease of remote warfare technology and lack of oversight
made armed drones experimental (Lindsay 2020; Sanger 2017). For example,
the U.S. was initially comfortable striking targets based on the
appearance of terrorist activity. This activity, which could consist of
younger men conducting jumping jacks at an apparent militant training
camp, circumvented positive identification of a high-value target based
on a confluence of intelligence (Sanger 2017). These so-called
``signature strikes'' were criticized for resulting in civilian
casualties because they relied on little more than observing a target's
pattern-of-life (Renic 2020). Over time, the utilization of these
strikes presented a moral hazard that incentivized the use of remote
warfare technology given the anticipated reduction of U.S. soldiers'
exposure to harm on the battlefield. The perceived merit of drone
strikes, however, came at a key cost: non-combatant immunity.

Indeed, this moral hazard encouraged the Obama administration to
authorize drone strikes in undeclared theaters of operations at ten
times the rate of the preceding Bush administration. This resulted in
the Obama administration's use of a drone strike every 5.4 days over the
course of its tenure (Zenko 2017). Obama argued ``the drones probably
had less collateral damage, which is the antiseptic way of saying it
killed people who were innocent and not just just targets.'' This was
troubling because the ``machinery of it started becoming too easy'' and
``turns out I'm really good at killing people'' (Zenko 2017).
Politically, in this era, drone strikes were broadly recognized as the
``least bad option'' to contain terrorists enjoying sanctuary in fragile
and failing states, including Pakistan, Somalia, and Yemen. These
operations gave the appearance of tough action against terrorists but
without the potential risks that accompany the deployment of U.S.
soldiers abroad, especially their redeployment home in body bags.

\hypertarget{near-certainty-of-risk-mitigation}{%
\subsection{``Near'' Certainty of Risk
Mitigation}\label{near-certainty-of-risk-mitigation}}

In response to the growing evidence of civilian harm from the U.S.
counterterrorism policy of drone strikes, a chorus of criticism emerged.
In 2010, the UN's Special Rapporteur on Extrajudicial, Summary, or
Arbitrary Executions, Philip Alston, released a report pointing to the
prolific use of drone strikes as the crux of counterterrorism efforts as
well as the implications for civilians. Alston implored the U.S. to
demonstrate more restraint when using armed drones, particularly in
undeclared theaters of war -- challenging the U.S. to clarify the legal
basis for its strikes in non-active areas of conflict (Alston 2010).
Other international and non-government organizations chimed in with
similar criticisms (Abizaid and Brooks 2015).

In interviews conducted at the end of his administration, Obama admitted
that pressure from these criticisms encouraged him to implement measures
for administrative officials to exercise more oversight on U.S. drone
strikes. He referenced the critiques of two non-governmental
organizations, the Human Rights Watch and Amnesty International, in
particular. ``So there's an example of where I think, even if the
criticism is not always perfectly informed and in some cases I would
deem unfair, just the noise, attention, fuss probably keeps powerful
officials or agencies on their toes. And they should be on their toes
when it comes to the use of deadly force'' (Friedersdorf 2016).

As civilian casualties and criticism mounted following U.S. drone
strikes in undeclared theaters of war, Obama directed a review of U.S.
counterterrorism policy during his first term in office. Obama
``directed his administration to tighten procedures and standards''
governing U.S. drone strikes abroad (Savage 2016). He reportedly
``pushed his staff to come up with a complex set of rules and legal
structures to make sure each strike comported to the rules of law''
(Sanger 2017). Although the near certainty standard was not publicly
disclosed until Obama's speech at the National Defense University in
late May 2013, this more restrictive guideline for civilian protection
during U.S. drone strikes was the result of intensive policy
deliberations starting as early as 2011 (Jaffer 2016; G. Martin and
Steuter 2017). Moyn characterizes the deliberations as designed to
achieve ``little more than self-regulation'' with the Obama
administration ``devising rules and enforcing them against themselves''
(Moyn 2021). Jaffer (2016) adds ``Obama administration officials
insisted that drone strikes were lawful, but the `law' they invoked was
their own.'' Even so, during the intervening period between 2011 and
2013, the Obama administration ``ratcheted up'' requirements for strike
approval in undeclared theaters of war to ensure more discriminatory
operations (Lindsay 2020).

Not only was the Obama administration's near certainty standard
important for instilling higher degrees of morality and legality into
the drone strike approval process, it was also designed to enhance
America's approbation by attenuating images of the ``quasi-secretive''
use of armed drones (Banka and Quinn 2018). By 2012, for instance, a Pew
Research Center poll showed that 94\% of Pakistanis thought that U.S.
drone strikes in Pakistan were killing ``too many'' civilians (Jaffer
2016). Public opinion tracked closely with the dominant narrative
portrayed in Pakistani newspapers from 2009-2013, which suggested that
U.S. drone strikes killed more civilians than combatants (Fair and Hamza
2016). In demonstrating the United States' rightful wartime conduct, in
this case using force to kill terrorists while minimizing risks to
innocent civilians, the near certainty standard was calibrated to
enhance foreign audiences' support for America's use of strikes abroad
(Aslam 2013; Clark 2005). Lastly, given the emerging arguments about
blowback -- that strikes with high civilian casualties might provide a
recruitment tool for more terrorists (Cronin 2020; Tirman 2011) -- the
U.S. had important instrumental reasons to minimize collateral damage.

\hypertarget{operationalizing-near-certainty}{%
\subsection{Operationalizing ``Near''
Certainty}\label{operationalizing-near-certainty}}

The genesis of the Obama administration's near certainty standard
suggests that any reduction in civilian casualties during U.S. drone
strikes in undeclared theaters of war is not a function of a figurative
piece of paper, the PPG. Rather, we anticipate a reduction of civilian
casualties based on the Obama administration's deliberative process of
carrying out a strike and the official policy change that publicly
acknowledged heightened scrutiny. Testing this hypothesis requires that
we first understand the mechanics of U.S. drone strikes to identify what
changed given the Obama administration's adoption of a more restrictive
targeting policy. Casting U.S. strikes in terms of a production function
suggests that the near certainty standard constituted the key mechanism
that changed aspects of the targeting process to further protect
civilians. This casts doubt on alternative explanations that suggest
conditions unique to Pakistan or shifts in the intended targets may
better account for the evolving accuracy of U.S. drone strikes (Plaw,
Fricker, and Williams 2011).

Drone strikes result from a two-stage process of intelligence and
approval that are linked by the certainty standard. Intelligence is the
process of analyzing raw data and information given some military or
political objective (Odom 2008). Approval is the prosecution of a strike
based on authorization from a political official. Intelligence drives
drone strikes and is a function of three actors. First, analysts assess
the body of reporting on a proposed target to justify its value -- the
target's placement, access, and contribution to a terrorist group.
Second, an intelligence officer communicates this assessment to a
commander while further validating the target's location, estimating the
impact of the target's loss, and framing the target against a
commander's priorities that reflect broader strategic and policy goals
(Lushenko 2018). Third, drone operators have the ability to shape
judgment for the degree to which strikes meet a certainty standard given
their prolonged observation of a target. Only in the case that targeted
killing is manifestly illegal, meaning it purposefully contradicts the
\textit{jus in bello} (justice in war) principle of distinction, will
operators protest a strike (M. Martin and Sasser 2010). Even then,
protests are usually \textit{post hoc} considering strikes are
ultimately underwritten by commanders.

Intelligence is decisive to the approval process that follows. With this
initial risk assessment, the approval process also includes two
additional assessments that further frame the certainty standard. A
collateral damage estimate (CDE) results from ``computerized algorithms
to predict, estimate, and minimize collateral damage'' based on the
munitions, terrain, urbanization, and human traffic pattern throughout
an area that is marked for a drone strike (Crawford 2013). The CDE
process further considers the potential for secondary explosions
following a drone strike that can -- and often do -- harm civilians
(Khan 2021). A lawyer then assesses the merits of targeted killing based
on the rules of engagement that link lethal action to a combatant's
hostile act or intent, as well as a politically-informed ``noncombatant
casualty cutoff value'' that shifts up or down based on the certainty
standard (Crawford 2013; Khan 2021; Liddick 2021).

Under the Bush administration, the United States' strategic use of drone
warfare was based on the lenient standard of reasonable certainty. This
standard allowed for civilian casualties in both declared and undeclared
theaters of war during U.S. drone strikes. In response to the concern
over the loss of civilian life, the Obama administration conditioned the
approval of drone strikes on four requirements. A strike would be
approved after demonstrating (1) a target constituted a ``continuing
imminent threat to U.S. persons''; (2) infeasibility of capturing the
target; (3) near certainty of target identification; and, (4) near
certainty of no civilian casualties\footnote{{``Remarks by the President
  at the National Defense University''} (2013).}.

Movement from reasonable to near certainty modified the intelligence and
approval processes in part by shifting the burden of proof from the
signature of terrorist behavior to positive identification of a target
as the threshold for a drone strike. Commanders also placed more
emphasis on the CDE assessment. This was crucial to ensure the Obama
administration's use of drone strikes for targeting killing aligned with
publicly acknowledged criteria and international law (B. DeRosa and
Regan 2021). Indeed, the near certainty standard resulted in a series of
military, legal, and policy checks. These required commanders to more
thoroughly inform ``decisions about whether a potential target satisfies
the policy criteria to be designated for a lethal use of force;
development of an operational plan, including assessment of whether it
is possible to satisfy operational requirements; legal review; final
policy-level approval, and some external oversight'' (B. DeRosa and
Regan 2021).

\hypertarget{theoretical-mechanism}{%
\subsection{Theoretical Mechanism}\label{theoretical-mechanism}}

We are interested in studying a change from the reasonable to near
certainty standard that came to govern the Obama administration's risk
assessment of using drone strikes in undeclared theaters of war,
including Pakistan. As described above, the production function for a
drone strike is split into two stages with separate actors. In studying
the change in certainty standard, we observe increased scrutiny for risk
assessment in several of these actors. The two stage process for a drone
strike's approval takes the following functional form:

\[StrikeApproval_s = f(Intelligence_s ; \quad CDE Assessment_s; \quad Legal Counsel_s)\]
\[ \text{where} \quad Intelligence_s = f(w_1*AnalystReporting; \quad w_2*IntelOfficer; \quad w_3*Operator)\]
We describe the function and role for each of these actors in our
conceptual framework above. We attach weights -- \(w_n\) -- to each of
the three actors in the intelligence gathering process and observe in
this setting that \(w_2 > w_1 > w_3\). The more stringent level of
certainty levied by the Obama administration's policy adjusted the risk
assessment guidelines to gain approval for a strike. Following the Obama
administration's implementation of the near certainty standard,
increased scrutiny was placed on three components of this process: the
intelligence officer's assessment, the CDE assessment, and legal
counsel.

\hypertarget{empirical-approach}{%
\section{Empirical Approach}\label{empirical-approach}}

\hypertarget{drone-strikes-data-and-outcomes}{%
\subsection{Drone Strikes Data and
Outcomes}\label{drone-strikes-data-and-outcomes}}

We use data from the BIJ that captures the universe of U.S. drone
strikes in Pakistan from 2002-2019. Specifically, these data capture 351
U.S. drone strikes. From these data, we retrieve strike-specific
information about location and estimates for civilian casualties, child
casualties, and total casualties. We opt to use data from the BIJ for a
number of reasons. First, the BIJ provides multiple references for each
strike cataloged in the dataset. The casualty values in the dataset are
validated by multiple sources that are made separately available and
when cross-referencing a subsample of the BIJ citations, we also
identified no reporting errors. Second, the U.S. government's estimates
for civilian casualties are often questioned by scholars and
practitioners. The author of the policy we study states this plainly:
``the U.S. figures undercount civilian casualties''\footnote{See the
  appendix for information on our elite interviews.}. Finally, the other
two organizations that similarly collect data on U.S. drone warfare --
New America Foundation and The Long War Journal -- are often criticized
for consistency issues (Kreps 2016). For our main analysis, we aggregate
our sample to the country-month level and observe strikes in 61 months
prior to the policy's implementation and 43 months afterwards.

We take the midpoint of the minimum and maximum estimates for casualties
to use as our outcome variables, following prior literature (Sheehan,
Marquardt, and Collins 2021). We construct these midpoint values for
civilian, child, and total casualties. We create a measure of strike
precision from these values as the proportion of total deaths from a
strike that are reported as combatant deaths. This measure of precision
ranges from 0 to 1 and takes the value of 1 when only combatants are
killed and 0 when only civilians are killed and can be interpreted as
the percent of total deaths from a strike that were combatants. We
aggregate these measures to the monthly-level for analysis.

\hypertarget{regression-discontinuity-in-time}{%
\subsection{Regression Discontinuity in
Time}\label{regression-discontinuity-in-time}}

Our baseline model is a regression discontinuity in time (RDiT) that
follows work by Davis (2008), Auffhammer and Kellogg (2011), Chen and
Whalley (2012), and Anderson (2014). This method mimics a randomized
controlled trial by observing the period before a policy goes into
effect as the counterfactual (control) to the treated observations. An
RDiT design is an extension of the canonical RD design to time series
data that treats calendar time as the running -- or forcing -- variable.
The discontinuity is estimated at a cutoff \(c\) that aligns with the
implementation of our policy. We opt for this design over our ideal
approach -- a difference-in-differences (DiD) model -- due to data
limitations in our only potentially valid counterfactual setting:
Afghanistan. We follow the guidance of Hausman and Rapson (2018) who
outline a set of assumptions and robustness checks that researchers
should satisfy when adopting this method.

\hypertarget{rd-specification}{%
\subsubsection{RD Specification}\label{rd-specification}}

Our primary regression discontinuity model takes the following form:

\[y_{m}=\alpha+\beta Cutoff_{m}+f\left({\mathrm{\ date\ }}_{m}\right)+\varepsilon_{m}\]

In this equation \(y_{m}\) is our monthly outcome variable in Pakistan
for month \(m\), \(Cutoff_{m}\) is a binary variable equal to unity when
the near certainty standard is implemented and zero otherwise, and
\(date_m\) is the unit measured in months from policy implementation. An
RD model is driven by the assumption that the potentially endogenous
relationship between our running variable \(date_m\) and the error term
\(\varepsilon_m\) is eliminated by the flexible function \(f(\cdot)\).
An assumption underlying this design is that the relationship between
the error term and our running variable does not change discontinuously
near our cutoff \(c\). In particular, the relationship between
\(\varepsilon_{m}\) and the date must not change discontinuously on or
near the date on which the strike begins. Our RD specification is a
sharp RD in that the running variable \(date_{m}\) completely determines
\(PPG_{m}\).

To estimate this model we follow Imbens and Lemieux (2008), using a
parametric approach with two bandwidths: the MSERD optimal and a manual
bandwidth of 48 months. We estimate local linear regressions of the
form:

\[ y_m = \alpha_0 + g(date_m)\beta + \alpha_1Cutoff_m + f(date_m)*Cutoff_m\omega + \epsilon_m \]

We estimate these regressions for outcomes \(y_m\) and create two
functional forms on either side of the discontinuity: \(f\) and \(g\).
We are interested in the coefficient \(\alpha_1\) which represents the
discontinuity in the trends between functions \(f\) and \(g\). Though we
capture strike-level outcomes in our aggregate measures, our primary
analysis is at the month-level. In contrast to studies of the military
effectiveness of individual strikes in Pakistan (P. Johnston and Sarbahi
2016) that encourage analysis at the day or week level, we opt for
monthly aggregation. This allows us to capture the broader impact for
U.S. counterterrorism strategy following a change in the certainty
standard -- particularly because strike frequency did not see an
analogous discontinuity in a parametric setting. Indeed, the Obama
administration intended the near certainty standard to radically alter
decision-making at the strike-level, thus reflecting a broader strategic
adjustment. Obama's articulation of the policy makes clear that the
intended effect was a net reduction in civilian casualties, invariant of
strike frequency\footnote{{``Remarks by the President at the National
  Defense University''} (2013).}.

We set our primary cutoff value \(c\) -- based on a series of analyses
and feedback from Obama-era officials responsible for implementing the
policy -- at July 2011. We perform an analogous specification for the
Obama administration's official announcement of the policy on May 23,
2013 -- though, we opted to not use this date for three key reasons.
First, we know that the certainty standard accompanying the Obama
administration's policy was being rolled out well before the official
policy announcement (Lindsay 2020). Indeed, the cutoff coincides with
several official statements and documents released by the Obama
administration relating to U.S. drone strikes, including a Justice
Department memorandum on the targeting of U.S. citizens abroad (Jaffer
2016). Second, setting the cutoff at the announcement date includes the
implementation period of the policy as untreated observations and
excludes several strikes with high civilian casualties -- notably in
July 2012. Third, we conduct a structural break analysis to corroborate
feedback from Obama-era officials indicating that July 2011 aligns with
the ``true'' policy implementation period.

\hypertarget{structural-break-analysis}{%
\subsubsection{Structural Break
Analysis}\label{structural-break-analysis}}

Our specification functions well when a discrete, sharp cutoff \(c\)
exists. Given this requirement, the roll-out period for the Obama
administration's policy of near certainty in the 18 to 30 months prior
to the official announcement could present an issue. To confirm July
2011 aligns with the ``true'' policy implementation date, as suggested
by our primary research, we conduct a structural break analysis.
Structural break analysis relies on Chow's F-test to estimate points in
time series data where there exists a pronounced break in the trend
underlying the outcomes. We use the {[}xtbreak{]} function in Stata to
estimate potential breaks in the BIJ data (Ditzen, Karavias, and
Westerlund 2021). Empirically, our structural break analysis takes the
following form:

\[
\hat{\mathcal{T}_{s}}=\arg \min _{\mathcal{T}_{s} \in \mathcal{T}_{s, \varepsilon}} S S R\left(\mathcal{T}_{s}\right)
\] The resulting estimate \(\hat{\mathcal{T}_{s}}\) provides the points
along our running variable of calendar time where the outcomes break
with the linear trend. We perform this estimation without hypothesized
breakpoints and instead specify three potential breaks. The estimate is
then evaluated on statistical power following Bai and Perron (1998). In
performing this operation, we find structural changes in our time series
that align across outcomes and aggregation level, though our monthly
breakpoints are better identified.

\hypertarget{results}{%
\section{Results}\label{results}}

\hypertarget{summary-statistics-and-trends}{%
\subsection{Summary Statistics and
Trends}\label{summary-statistics-and-trends}}

\hypertarget{monthly-summary-statistics-for-strike-outcomes}{%
\subsubsection{Monthly Summary Statistics for Strike
Outcomes}\label{monthly-summary-statistics-for-strike-outcomes}}

\bgroup
\def\arraystretch{1.5}

\begin{table}[h]
\centering
\caption{Summary Statistics for Casualty Outcomes}

\begin{tabular}{lcccc}
\hline
 & [Pre Jul11] & [Post Jul11] & [Pre May13] & [Post May13]\\
\hline

\bf{Civilian Casualties} & ~ & ~ & ~ & ~\\

~~ Count & 607 & 90 & 690 & 7\\

~~ Mean & 11.902 & 1.689 & 9.318 & 0.233\\

~~ Mean MinEst & 7.667 & 0.623 & 5.689 & 0.1\\

~~ Mean MaxEst & 16.137 & 2.755 & 12.946 & 0.367\\

\bf{Child Casualties} & ~ & ~ & ~ & ~\\

~~ Count & 182 & 7 & 188 & 1\\

~~ Mean & 3.578 & 0.132 & 2.547 & 0.033\\

~~ Mean MinEst & 3.294 & 0.075 & 2.324 & 0\\

~~ Mean MaxEst & 3.863 & 0.189 & 2.77 & 0.067\\

\bf{Total Casualties} & ~ & ~ & ~ & ~\\

~~ Count & 2256 & 1014 & 2936 & 334\\

~~ Mean & 44.235 & 19.142 & 39.676 & 11.15\\

\bf{Strike Precision} & ~ & ~ & ~ & ~\\

~~ Mean & 0.686 & 0.95 & 0.761 & 0.97\\

\bf{Strike Frequency} & ~ & ~ & ~ & ~\\

~~ Count & 263 & 167 & 369 & 61\\

~~ Mean & 5.157 & 3.151 & 4.986 & 2.033\\

\noalign{\vskip 8mm}

\end{tabular}

\textit{Table of means of primary outcome variables indexed by two cutoffs. July 2011 -- our policy cutoff -- and May 2013, when the Obama administration announced the policy. Mean values of the minimum and maximum estimates of casualties are also provided, from the BIJ, and shown for child and civilian deaths. Strike precision is the proportion of those killed from drone strikes that were the intended target.}

\end{table}

\egroup

\FloatBarrier

\hypertarget{trends-in-drone-strike-casualties}{%
\subsubsection{Trends in Drone Strike
Casualties}\label{trends-in-drone-strike-casualties}}

\hypertarget{cumulative-deaths-from-u.s.-strikes}{%
\paragraph{Cumulative Deaths from U.S.
Strikes}\label{cumulative-deaths-from-u.s.-strikes}}

We first visualize the arc of the U.S. drone program and its cumulative
human toll in Pakistan. We visualize cumulative deaths over time and
plot them below. We observe a period from 2005 to 2008 -- during the
Bush administration -- where civilian deaths in Pakistan accounted for a
majority if not the totality of casualties as a result of U.S. drone
strikes in that country. This poor targeting and precision of strikes
improves during the transition to the Obama administration -- denoted
with blue shading in the plots below. We observe an increase in
combatant casualties in tandem with a growth in civilian casualties
through the initial years of the Obama administration. This aligns with
the endogeneity of operational ability in the certainty standard that
governs civilian risk: strikes conducted with improved technology and
capabilities are still liable to harm civilians so long as they are
governed by the standard of reasonable certainty.

\FloatBarrier

\begin{figure}[ht]

\begin{center}\includegraphics[width=0.98\linewidth]{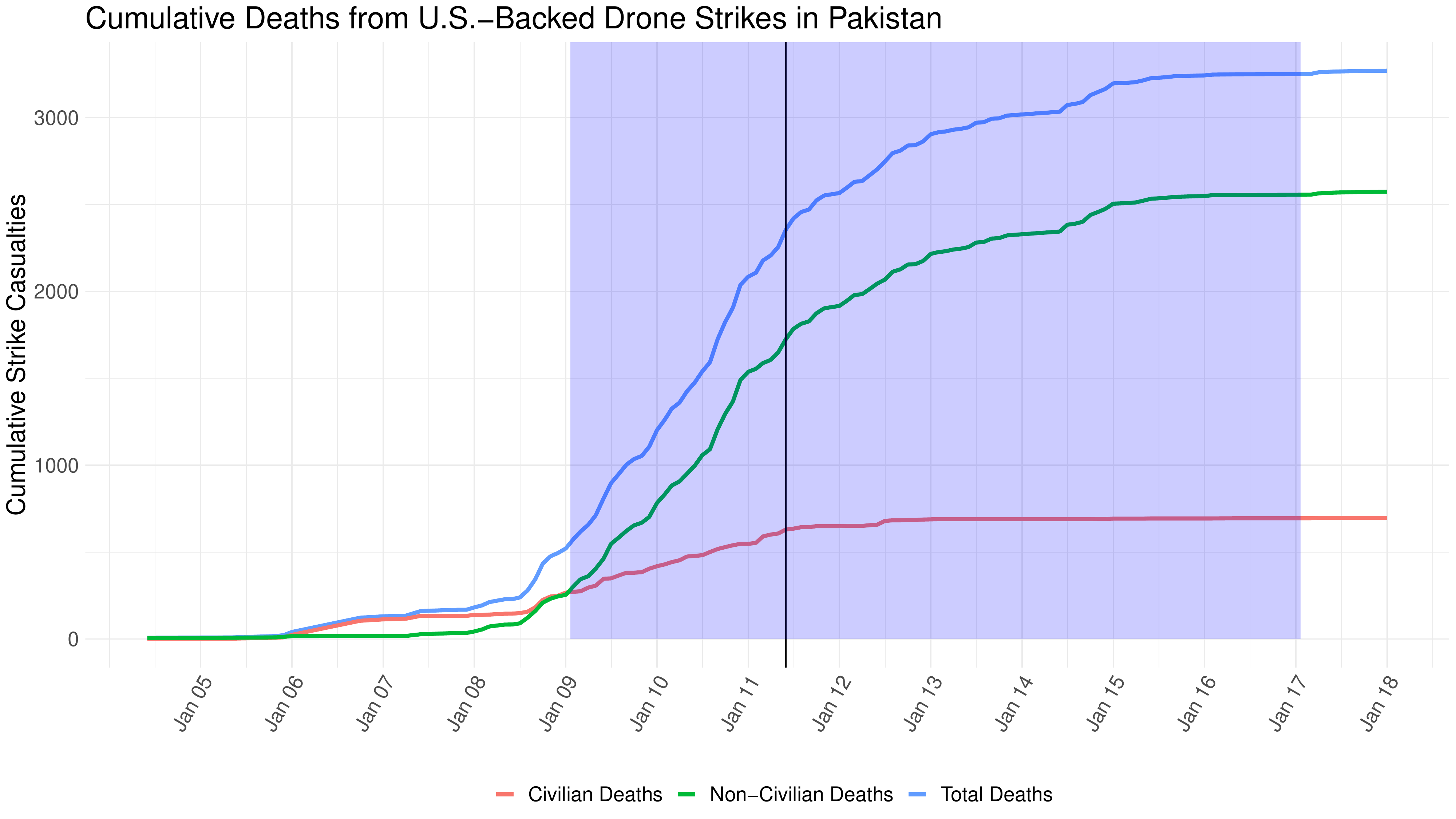} \end{center}

\begin{center}\includegraphics[width=0.49\linewidth]{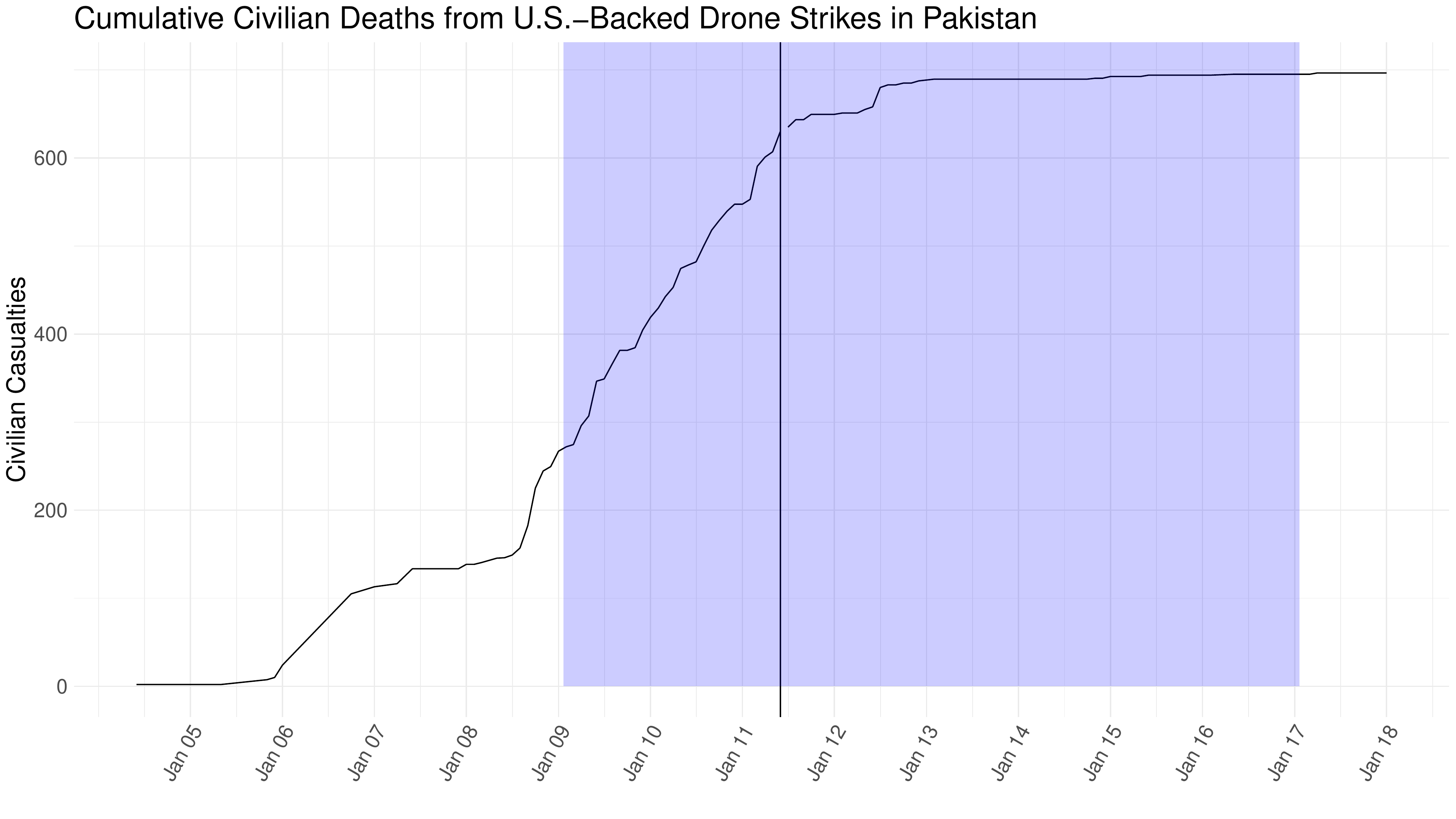} \includegraphics[width=0.49\linewidth]{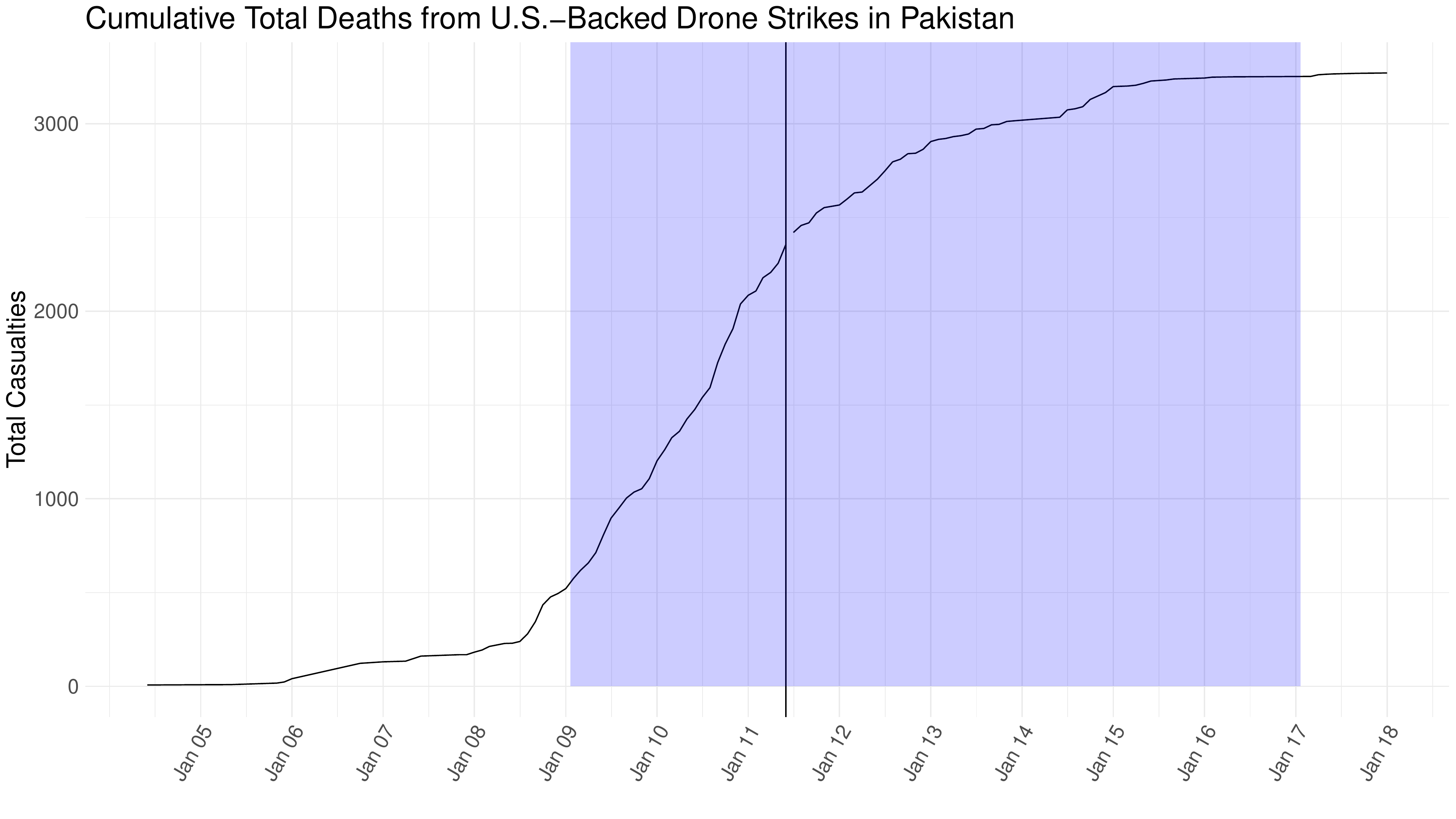} \end{center}

\caption{Plots display cumulative total, civilian, and combatant casualties from 2002-2019. Data retrieved from the Bureau of Investigative Journalism (BIJ). Plotted values represent the midpoint of minimum and maximum casualty estimates provided in the BIJ data. Shading reflects the Obama administration.}

\end{figure}

\hypertarget{civilian-casualties}{%
\paragraph{Civilian Casualties}\label{civilian-casualties}}

We study the shift to a near certainty standard that conditioned strike
approval on assurance of no civilian harm. We plot civilian casualties
resulting from strikes below and provide them at the strike- and
monthly-level -- we opt for the latter level of aggregation in our
primary RD specification. The vertical line in these plots denotes our
policy cutoff: July 2011. Mean civilian casualties are superimposed on
either side of the policy cutoff in red lines.

\FloatBarrier

\begin{figure}[ht]

\begin{center}\includegraphics[width=0.49\linewidth]{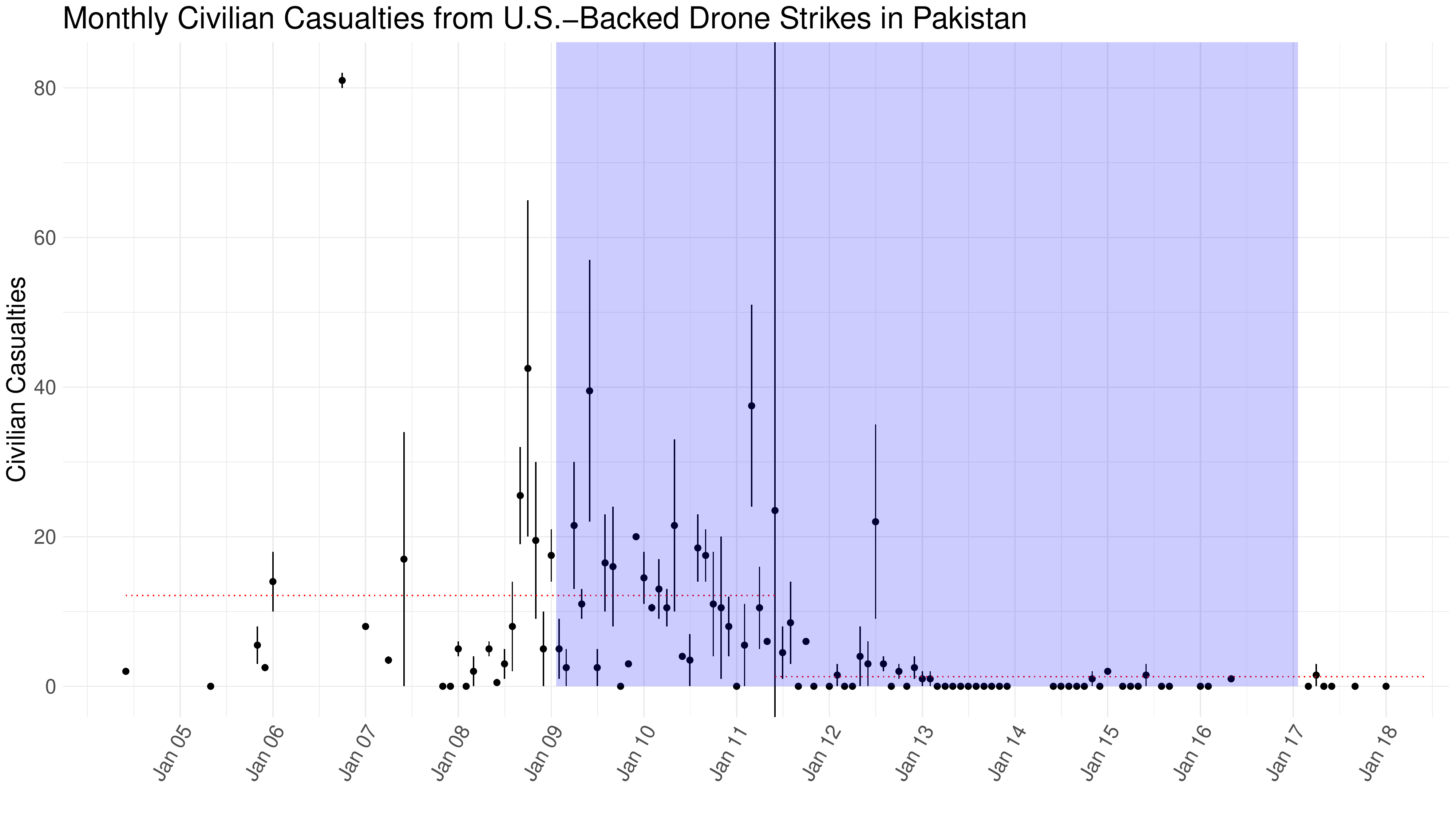} \includegraphics[width=0.49\linewidth]{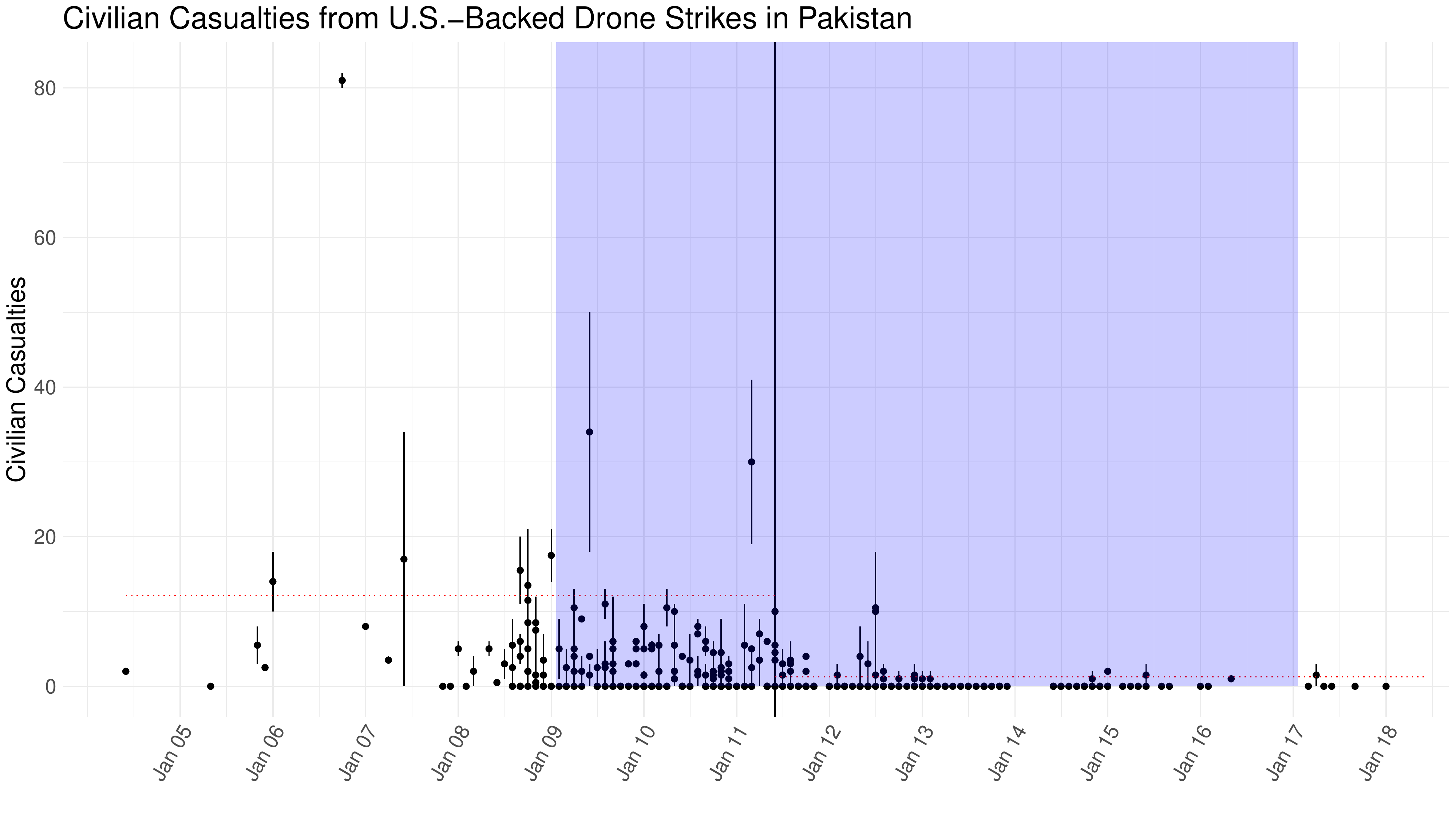} \end{center}

\caption{Data sourced from the Bureau of Investigative Journalism. Observations are midpoint of casualty estimates and error bars are min/max estimates.The superimposed dotted lines on either side of July 2011 represent the outcome variable (y axis) mean for the period before and after near certainty standard implementation. Shading reflects the Obama administration.}

\end{figure}

We observe that the mean civilian casualty value reduces to almost zero
at both the strike- and monthly-level after implementation of the near
certainty standard. This reduction in civilian deaths drives our RD
specification and creates a discontinuity at the implementation of the
policy.

\hypertarget{strike-precision}{%
\paragraph{Strike Precision}\label{strike-precision}}

The near certainty standard was put in place to mitigate civilian harm
though scholars argue that the observed reductions may be more
attributable to fewer strikes rather than an increase in strike-level
precision (Plaw, Fricker, and Williams 2011). To the extent scholars do
study the implications of the Obama administration's near certainty
standard for targeting accuracy, they merely rely on descriptive trends
-- ``fewer injuries implies more decisive lethal operations'' (Lindsay
2020). Rather, we construct a measure of strike precision as the
proportion of non-civilian deaths among total deaths from a given
strike. This measure should take -- based on principles of
\textit{jus in bello} -- a value of 1 for every strike meaning that no
civilians were at risk and only targeted individuals were killed. We
illustrate presidential administrations and means analogous to the plots
of civilian deaths.

\FloatBarrier

\begin{figure}[ht]

\begin{center}\includegraphics[width=0.49\linewidth]{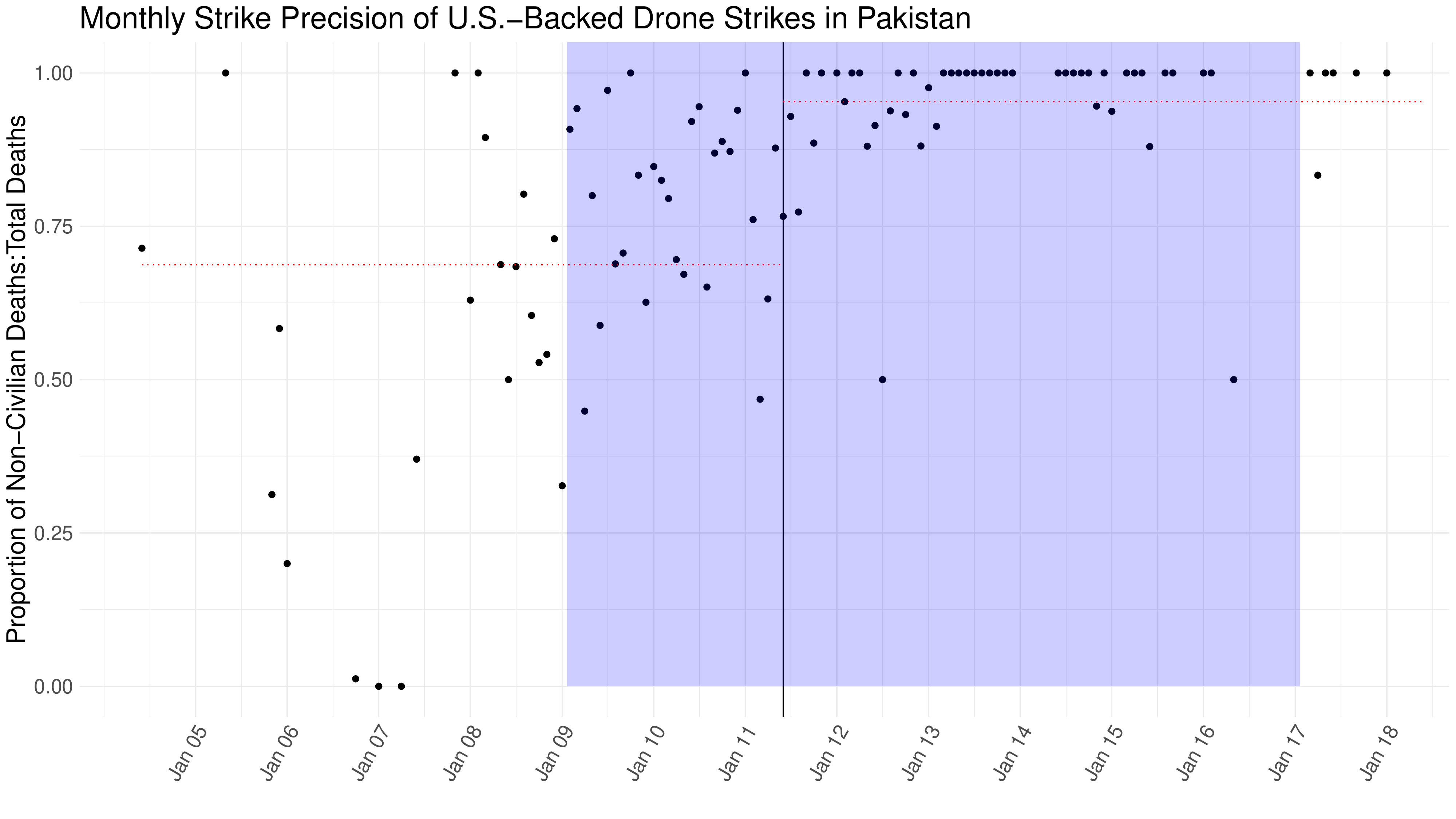} \includegraphics[width=0.49\linewidth]{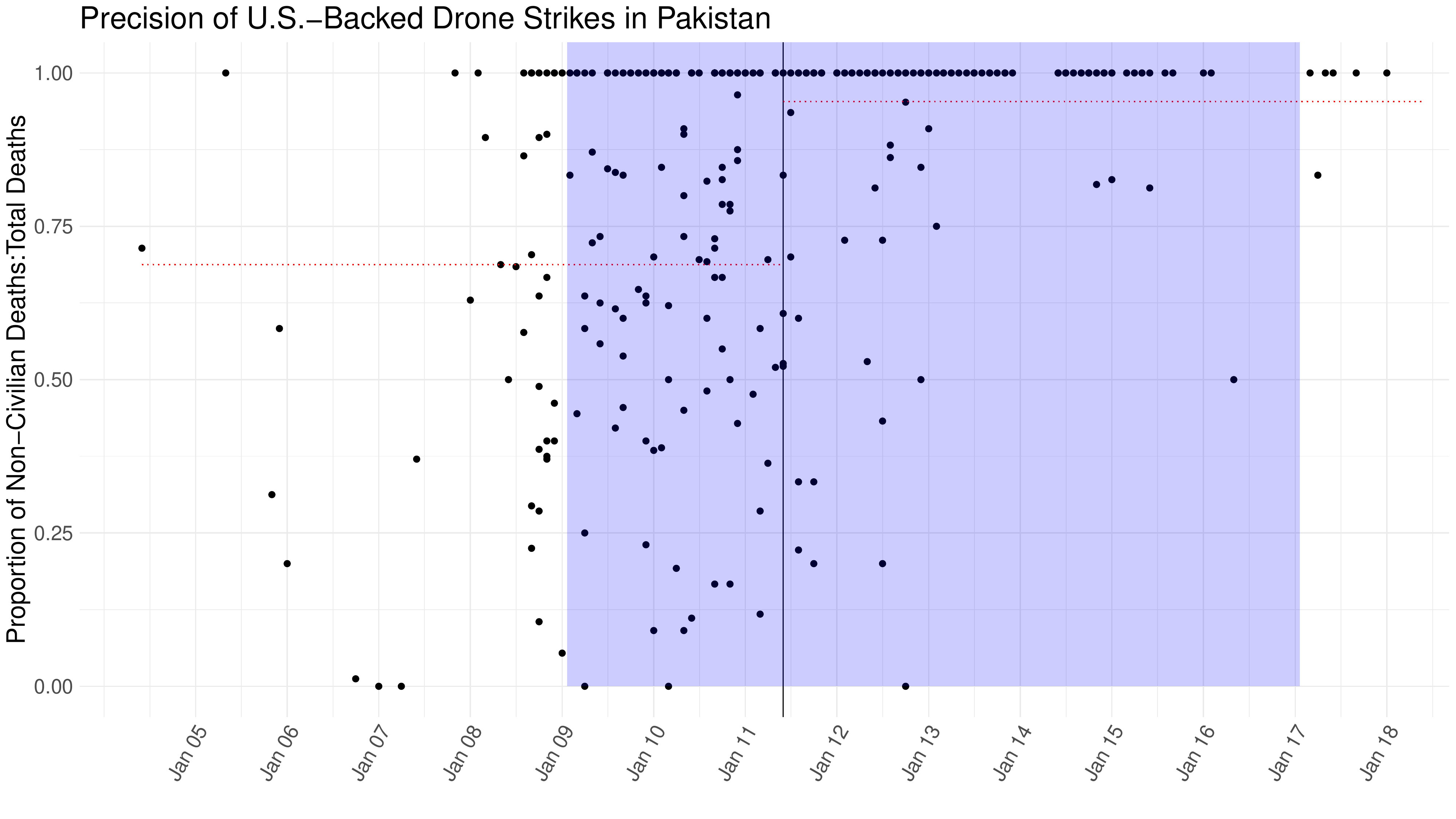} \end{center}

\caption{Data sourced from the Bureau of Investigative Journalism. Observations are proportion of those killed who were non-civilians (enemy combatants who were targeted by the strike). The superimposed dotted lines on either side of July 2011 represent the outcome variable (y axis) mean for the period before and after near certainty standard implementation. Shading reflects the Obama administration.}

\end{figure}

We observe the pre-period mean strike precision to be 0.686 -- less than
70 percent of those killed in strikes before the near certainty standard
were the intended target. The mean value of strike precision following
the Obama administration's adoption of the near certainty standard grows
to 0.95 or 95 percent precision in successful targeting.

\hypertarget{regression-discontinuity-results}{%
\subsection{Regression Discontinuity
Results}\label{regression-discontinuity-results}}

We estimate regression discontinuity models for the following outcomes:
(1) civilian casualties, (2) strike precision, and (3) civilian
casualties per strike. Table 2 presents RD estimates for monthly-level
outcomes using the policy cutoff from our structural break analysis --
July 2011 -- and two bandwidths: the MSERD optimal and a manual
bandwidth of four years.

\begin{table}[h]
\centering
\caption{Main Regression Discontinuity Estimates}

\begin{tabular}{lcccccc}
\toprule
  & Civ Cas & Civ Cas  & Str Prec & Str Prec  & Civ P/S & Civ P/S \\
\midrule
Conventional & \num{-11.139}+ & \num{-9.523}* & \num{0.177}+ & \num{0.089} & \num{-1.983}+ & \num{-0.861}\\
 & (\num{0.058}) & (\num{0.015}) & (\num{0.066}) & (\num{0.117}) & (\num{0.070}) & (\num{0.144})\\
Bias-Corrected & \num{-12.564}* & \num{-8.526}* & \num{0.196}* & \num{0.119}* & \num{-2.160}* & \num{-1.074}+\\
 & (\num{0.032}) & (\num{0.030}) & (\num{0.041}) & (\num{0.036}) & (\num{0.049}) & (\num{0.068})\\
Robust & \num{-12.564}+ & \num{-8.526} & \num{0.196}+ & \num{0.119} & \num{-2.160}+ & \num{-1.074}\\
 & (\num{0.069}) & (\num{0.144}) & (\num{0.088}) & (\num{0.147}) & (\num{0.100}) & (\num{0.230})\\
\midrule
PolyOrder & \num{1.000} & \num{1.000} & \num{1.000} & \num{1.000} & \num{1.000} & \num{1.000}\\
OrderBias & \num{2.000} & \num{2.000} & \num{2.000} & \num{2.000} & \num{2.000} & \num{2.000}\\
Kernel & Triangular & Triangular & Triangular & Triangular & Triangular & Triangular\\
Bandwidth & 22 months & 48 months & 17 months & 48 months & 16 months & 48 months\\
BWType & mserd & Manual & mserd & Manual & mserd & Manual\\
\bottomrule
\multicolumn{7}{l}{\rule{0pt}{1em}+ p $<$ 0.1, * p $<$ 0.05, ** p $<$ 0.01, *** p $<$ 0.001}\\

\end{tabular}

\bigskip

\textit{Regression discontinuity estimates are provided for three outcomes: monthly civilian casualties, strike precision defined as the proportion of casualties from a strike that were intended targets, and civilian casualties per strike. The cutoff date is set to July 2011. These are linear, parametric models following the specification detailed in Section 3.2.1 that are fit with an MSERD optimal bandwidth in odd numbered columns and a bandwidth of 4 years in others.}

\end{table}

\FloatBarrier

There is a statistically significant discontinuity at this cutoff for
all three of our outcome variables. We estimate a discontinuity in
civilian casualties that results in a 12 death reduction per month. It
is important to note that this aggregate measure of monthly civilian
casualties may reflect changes in strike frequency. Our model estimates
a reduction of 2 civilian casualties per strike following the
implementation of this higher certainty standard. This estimated
reduction is similar in magnitude to the average civilian casualty
estimate \textit{before} the policy was implemented. We augment this
further in our Monte Carlo simulation, below. Finally, we estimate an
increase in strike precision of 19.6 percentage points. The mean value
of this measure before the certainty standard had changed was 0.82 --
providing context to our estimate that we approach 100\% strike
precision following this change. We find no measurable discontinuity in
strike level outcomes and discuss this further in our robustness checks
below.

\hypertarget{robustness-checks}{%
\subsection{Robustness Checks}\label{robustness-checks}}

Our primary analysis is an extension of the canonical RD design. We
employ a series of robustness checks to verify the validity of the RD
design with emphasis placed on additional concerns we face using
calendar time as our running variable. We turn to guidance by Hausman
and Rapson (2018) for a checklist of robustness checks to validate the
assumptions of our RDiT specification. We attempt to address each of
their checklist items and provide justification in situations where we
are unable to. We present these checks below.

\hypertarget{plot-raw-data}{%
\paragraph{(1) Plot Raw Data}\label{plot-raw-data}}

Regression discontinuity designs rely on a change in the behavior of
data around some cutoff \(c\). In an RDiT specification, the behavior of
observations across time (trends) also provide an important context for
the time before a policy was implemented. This is especially important
as we consider bandwidth and cutoff selection. As such, we plot the raw
values of our outcome (civilian casualties) across time and fit linear
trends on either side of our chosen cutoff below:

\FloatBarrier

\begin{figure}[ht]

\begin{center}\includegraphics[width=0.49\linewidth]{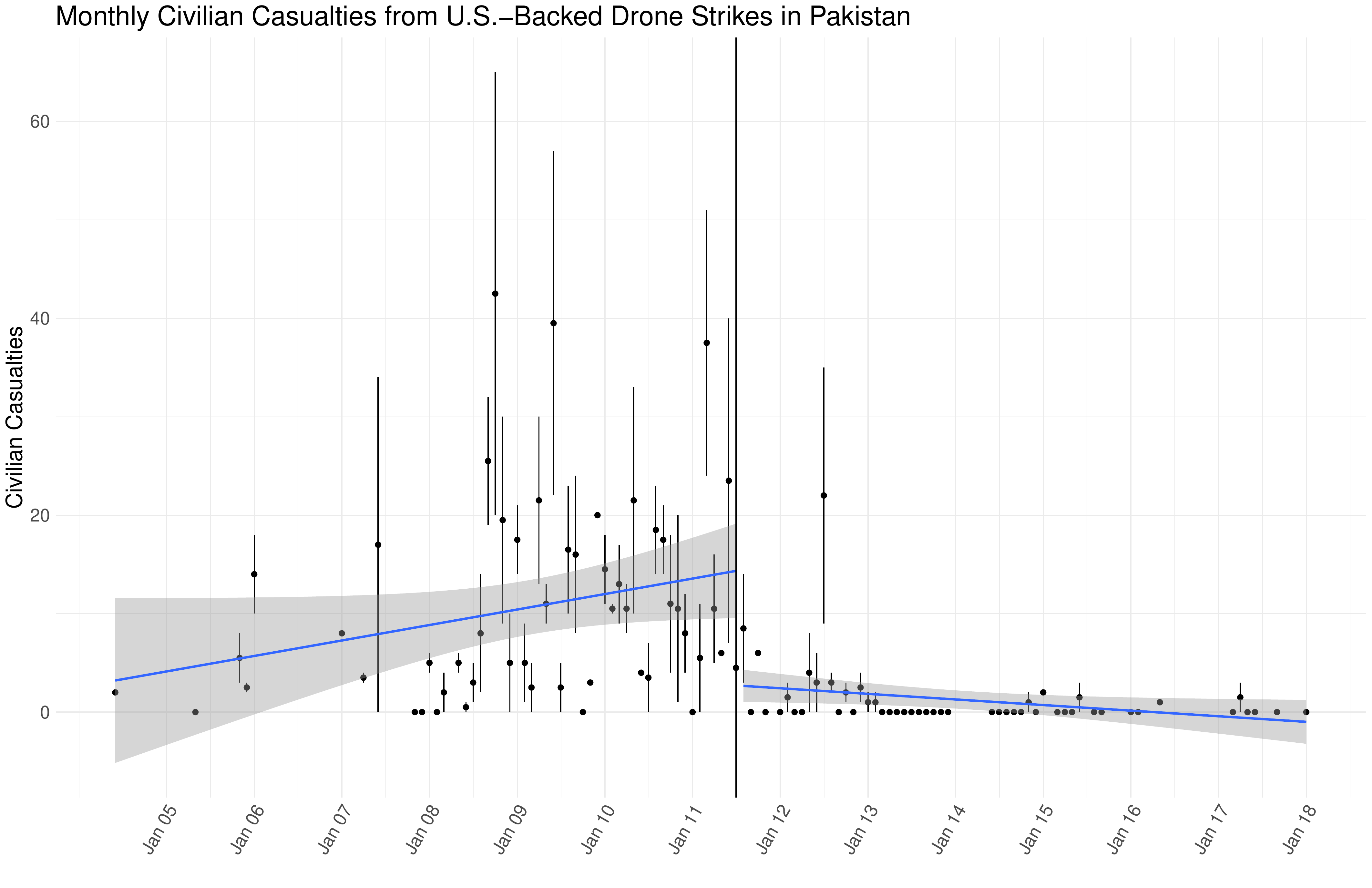} \includegraphics[width=0.49\linewidth]{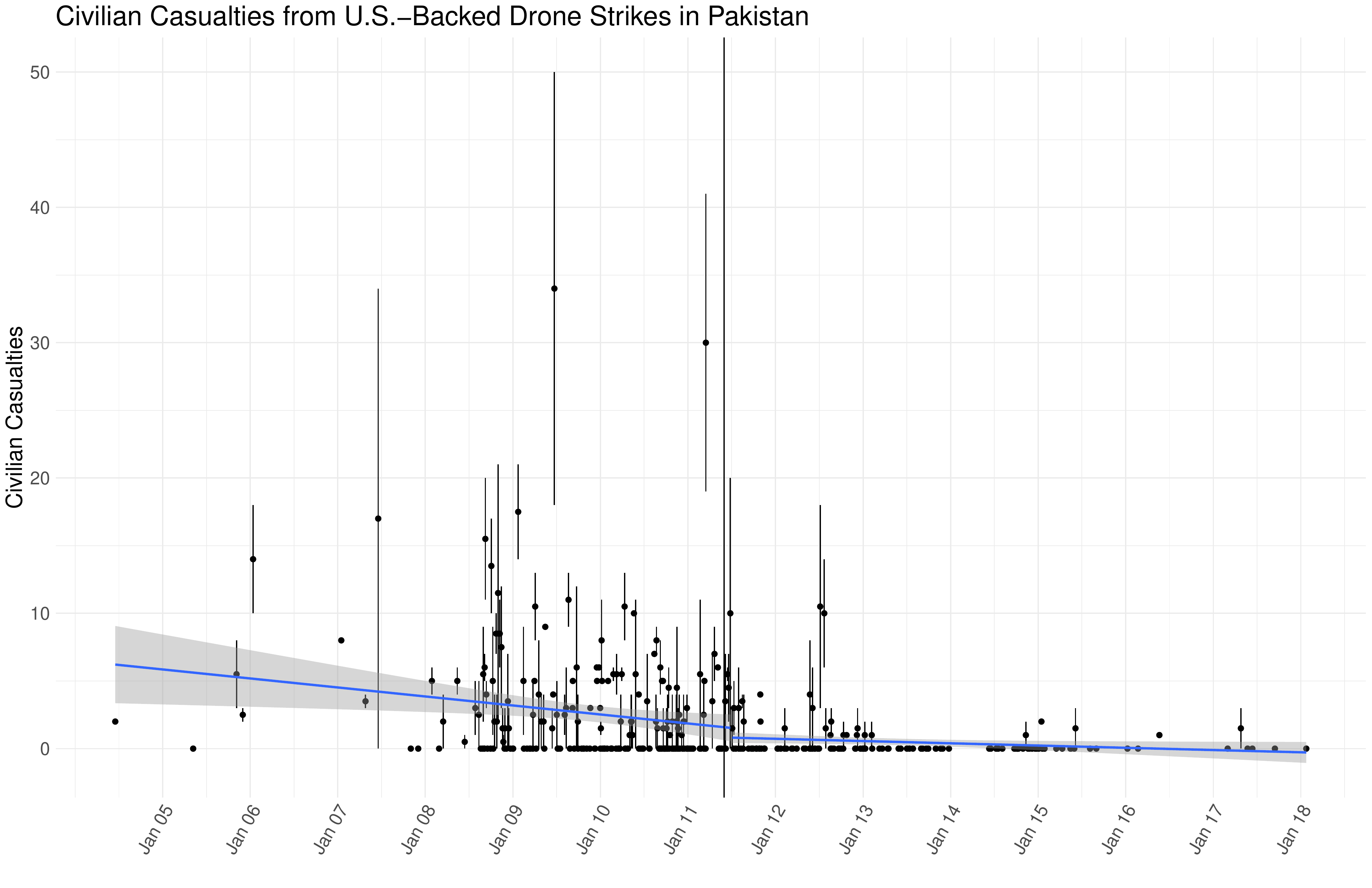} \end{center}

\caption{One observation removed for visualization [Oct 2006, 81 Deaths]. Data sourced from the Bureau of Investigative Journalism. Observations are midpoint of casualty estimates and error bars are min/max estimates. Vertical line drawn at July 2011 - the point of structural change in the trend of our outcome. Linear fit smoother applied for each side of this cutoff.}

\end{figure}

In these plots, the circular dots are midpoints of civilian casualty
estimates. The whiskers extending from each circular dot represent the
minimum and maximum estimated civilian casualties for that observation,
as provided by the BIJ. The cutoff \(c\) is represented by the vertical
line placed at July {[}1{]}, 2011 -- the estimated breakpoint in the
running variable from our structural break analysis that corresponds to
information gathered during our interviews with senior Obama-era
officials.

We do not observe a significant discontinuity for strike-level civilian
casualties. This aligns with the broader goal of this policy --
minimizing civilian casualties invariant of strike frequency by
implementing higher scrutiny at the strike-level. While a linear
discontinuity does not exist at the cutoff for strike-level casualties,
our ANOVA results support a measurable, statistically significant
difference in our outcome conditional on exposure to the near certainty
standard. Additionally, we do not see a discontinuity in monthly strike
frequency which bolsters our assertion that the mechanism that produced
monthly reductions occurred at the strike-level.

\hypertarget{parallel-rd-specifications}{%
\paragraph{(2) Parallel RD
Specifications}\label{parallel-rd-specifications}}

We fit identical, ``parallel,'' regression discontinuity models to other
outcomes within our data. We are interested in empirically testing if
the mechanism at cutoff \(c\) creates discontinuities in outcomes
outside of our framework. If we see a discontinuity in these other
outcomes, we have reason to doubt that the effect of our policy
mechanism is isolated to civilian casualties. We perform a series of
three parallel RD specifications. For outcomes we use: (1) strike
frequency, (2) combatant casualties, and (3) weather. We include these
three as they fall outside the scope of the policy and should not be
affected in a way that creates the discontinuity we observe in our
civilian casualty outcomes. We find no measurable, statistically
significant discontinuities in any of these falsification models.

\hypertarget{rdit-calibration}{%
\paragraph{(3) RDiT Calibration}\label{rdit-calibration}}

In addition to the structural break analysis, we further validate our
chosen cutoff by running a ``rolling'' RD specification. We conduct this
test to inspect the RD estimates at different cutoffs \(c\) by
iteratively changing \(c\) for every month in the two and a half years
prior to the Obama administration's official announcement of the near
certainty standard. This specification iterates \(c\) monthly from
October 2010 to May 2013. We plot the distribution of these estimates
across cutoff values to demonstrate the alignment of the most
significant RD estimate with the confidence interval from our structural
break analysis. This analysis corroborates our assertion that the policy
implementation date advanced by Obama-era officials aligns with a
discontinuity at our chosen cutoff. To determine how removing the
implementation period impacts our RDiT specification, which addresses
the issue of selection into treatment, we conduct a donut RD below.

\FloatBarrier

\begin{figure}[ht]

\includegraphics[width=\textwidth,height=3.64583in]{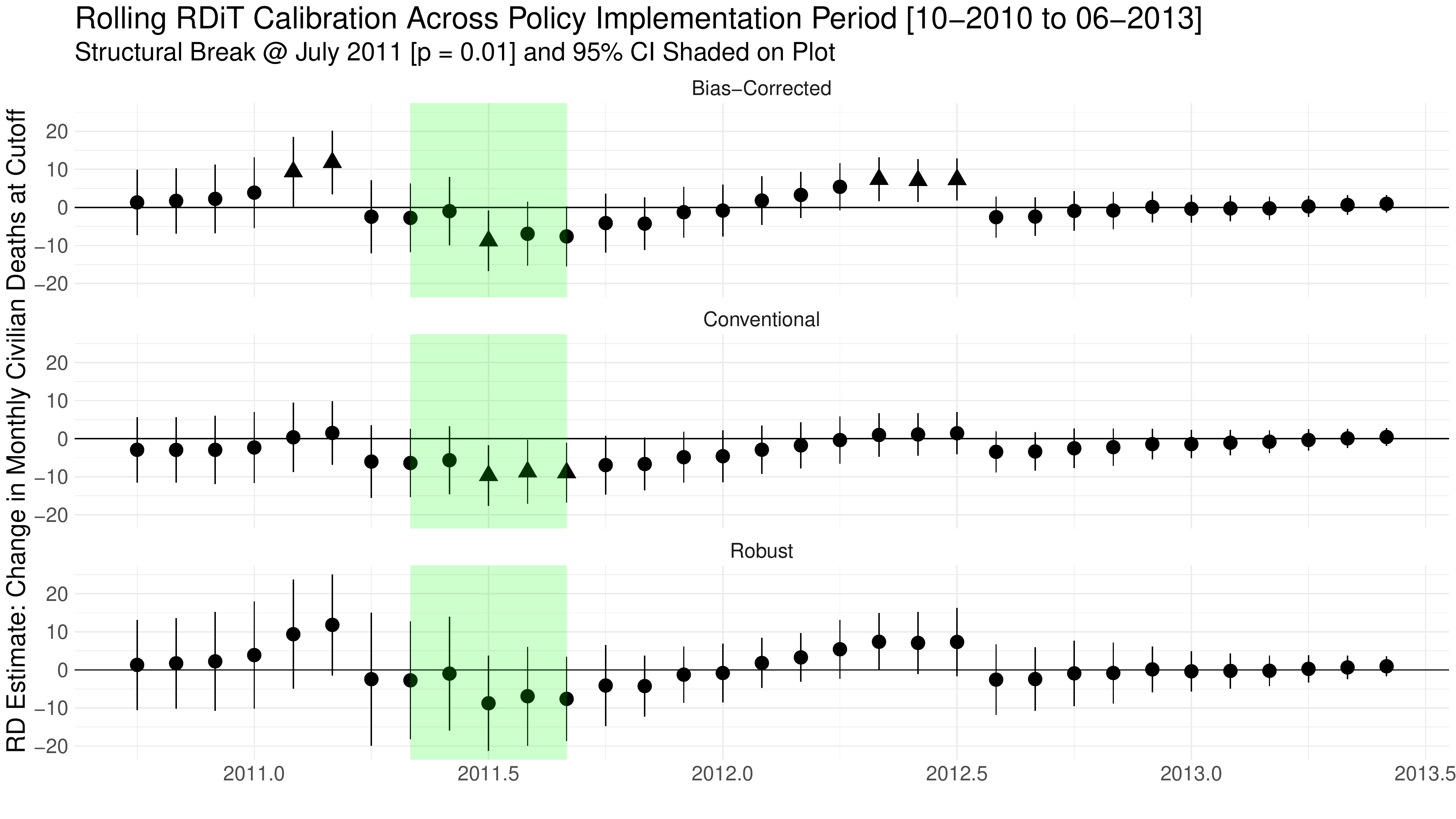}

\caption{RDiT specification "rolling" across policy implementation period: October 2010 to May 2013. Shading reflects confidence interval from structural break analysis: May 2011 to August 2011. Triangular points on the graph are statistically significant at the $\alpha = 0.05$ level. All models used a 48 month (4 year) bandwidth, triangular kernel, and a first-order polynomial (parametric linear RD).}

\end{figure}

\FloatBarrier

\hypertarget{donut-rd-specification}{%
\paragraph{(4) ``Donut'' RD
Specification}\label{donut-rd-specification}}

A ``donut'' regression discontinuity specification is a robustness check
to mitigate concerns about short-run selection, anticipation, or
avoidance effects. This specification is particularly useful in
scenarios where observations could potentially ``sort'' across the
cutoff to align with their preferences -- it is performed by removing
observations in a specified event window (the donut). In our setting,
being observed in the data at all is random and we would not expect any
potential sorting by way of this randomness. Instead, we adopt a donut
RD specification to evaluate differential effects across the
implementation period of Obama's near certainty standard. Our primary
cutoff \(c\) aligns with a structural break in our analysis, July 2011,
which comports with the initial rollout of the policy. Obama's public
announcement of this policy -- after being broadly implemented --
provides a separate cutoff on May 23, 2013. We identify this window
between 2011 and 2013 as the implementation period for this policy and
conduct an analysis where the largest donut we test spans this entire
distance. We divide this initial window into four donut specifications,
centered around April 2012 and expanding in diameter by roughly 6 months
until we estimate our full donut RD.

We estimate a donut RD design on both strike- and monthly-level civilian
casualties, analogous to the setting from our raw data plots. In the
plots below, the shaded area is our full donut that estimates the
discontinuity with the observations that remain on either end of the
cutoff. The donut RD plot for monthly civilian casualties is provided
below:

\FloatBarrier

\begin{figure}[ht]

\includegraphics[width=\textwidth,height=3.64583in]{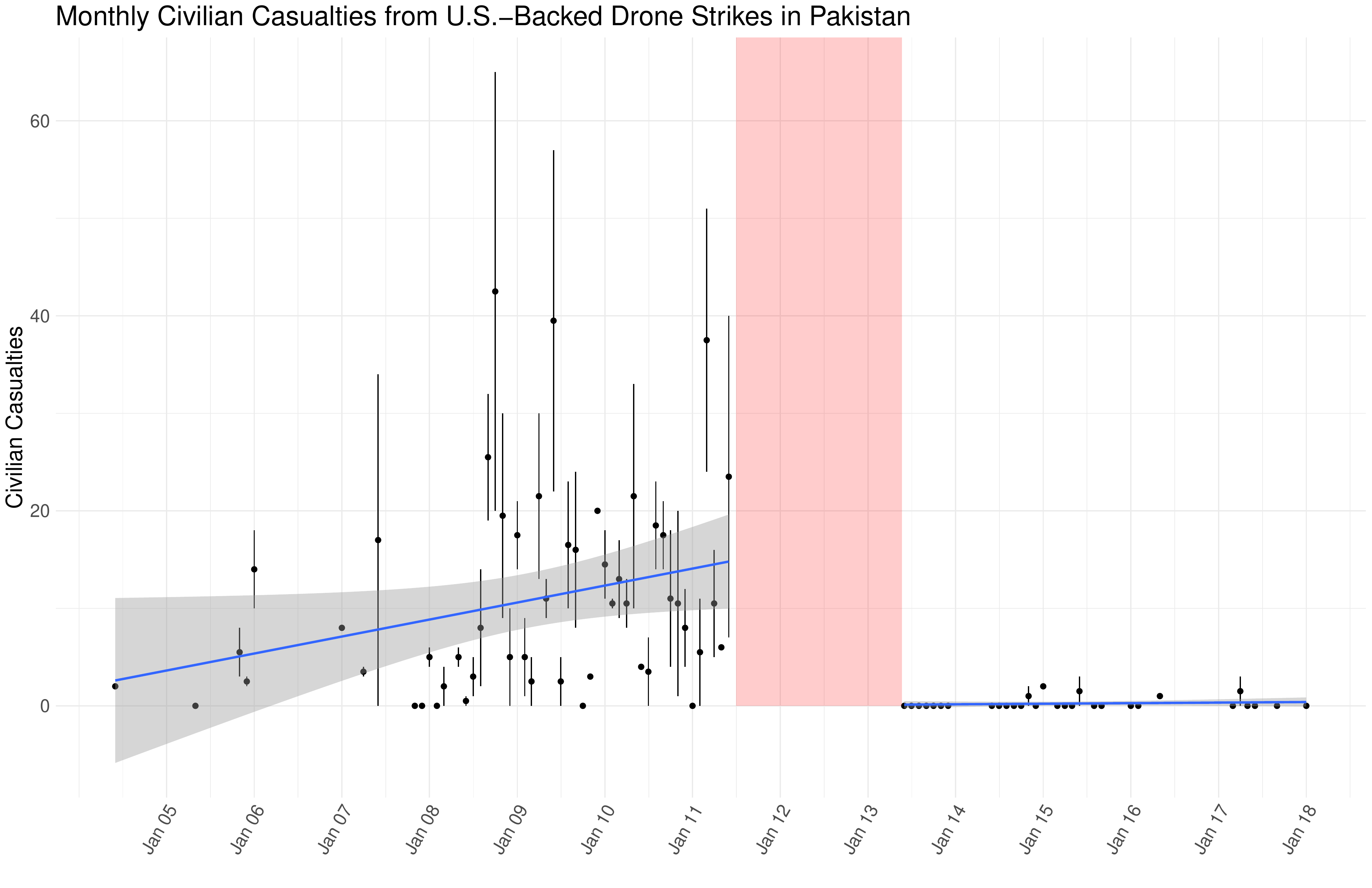}

\caption{Donut RD specification illustrated with the time between our chosen policy cutoff (July 2011) and the announcement of the PPG policy (May 2013) shaded out as the "donut" in this hypothesized model. Data sourced from the Bureau of Investigative Journalism. Observations are midpoint of casualty estimates and error bars are min/max estimates.}

\end{figure}

The discontinuity in this specification, shown in raw values above,
provides a stronger estimate than our standard RDiT (a reduction of
\textasciitilde15 civilian casualties per month). There are a number of
explanations for this finding. Importantly, it results from removal of
the policy rollout period that includes strikes resulting in nontrivial
civilian casualties during two separate operations in July 2012. Given
our data scarcity and the importance of including these two strikes, we
offer this donut specification only to reflect the effect of this policy
outside of its implementation period. We favor our primary specification
based on its inclusion of all our data.

\hypertarget{other-checks}{%
\paragraph{(5) Other Checks}\label{other-checks}}

The checklist also suggests providing several additional robustness
checks, including polynomial overfitting, geographic placebo parallel
RDs, serial autoregression, and an augmented local linear methodology
posed by the authors. We fit a linear trend at our discontinuity and
provide evidence of robustness to varying polynomial order -- though our
local linear regression estimates are strongest (Pei et al.~2020). We
are unable to perform parallel RD specifications in placebo geographies
due to the same data limitations that prevent us from implementing a DiD
design. We test for serial autoregression and find no motivation for a
lagged dependent variable specification. Autoregression in our time
series would suggest that the civilian casualties in time \(t\) would be
predictive of casualties in time \(t+n\) where
\(n \in 1, \cdots, t_{k}\). This is not the case in this setting as
civilian casualties represent, in their existence, a deviation from
military strategy and goals. The expectation for our outcome of civilian
casualties should be zero and are viewed as-if random otherwise
(Rigterink 2021). The final recommendation is the use of the Hausman and
Rapson (2018) augmented local linear methodology. The authors argue that
this procedure conditions on important regressors and can provide a more
effective estimate on a smaller bandwidth. Our setting and specification
provide strong estimates with a linear fit on either side of our cutoff,
invariant of covariate inclusion.

\hypertarget{discussion}{%
\section{Discussion}\label{discussion}}

\hypertarget{key-assumptions-limitations}{%
\subsection{Key Assumptions \&
Limitations}\label{key-assumptions-limitations}}

Our results rely on three key assumptions. First, we assume that
civilian casualties resulting from U.S. drone strikes are as-if random
-- it is impossible to predict civilian deaths during future strikes
(Rigterink 2021). Indeed, a recent report claims drone strikes ``hit
their targets with near-unerring accuracy'' (Khan 2021). This assumption
aligns with the just war component of \textit{jus in bello} and its
principle of distinction that implies civilian casualties result from
human error or the policy for risk mitigation (Crawford 2013). Second,
we contrast existing explanations of these reductions in civilian risk
that emphasize improvements in drone technology (Plaw, Fricker, and
Williams 2011) and posit that the near certainty standard is the primary
mechanism for the reduction of civilian casualties in Pakistan. While
technological advances may have increased the operational capacity of
drone warfare, they are endogenous to the near certainty standard that
governs the risk faced by civilians.

Finally, our policy implementation date is technically unknown. As such,
we choose our sharp cutoff \(c\) based on primary research and validate
it with myriad robustness checks. This estimated cutoff is corroborated
by the known, lagged announcement of the near certainty standard and the
swift, full-compliance nature of military guidance. Guidance for the
near certainty standard was being disseminated in the 18 to 30 months
prior to the May 23, 2013 announcement and the structural break analysis
identified a breakpoint in this range, as did our ``rolling'' RDiT
method. Additionally, a primary role of military officials is to rapidly
disseminate policy guidance through the chain-of-command (Owens 2011).
This swift implementation of military policy lends itself to our sharp
RD design -- one which relies on the cutoff to discreetly identify a
discontinuity.

This study is also vulnerable to limitations outside the scope of our
econometric specification. The intended targets in Pakistan may have
adjusted their behavior in ways that created more patience in military
strategy and reduced civilian casualties due to human error. Drone
strikes in Pakistan increased commensurately with the Obama
administration's surge of forces to Afghanistan -- in part as an
additional line of defense (Ullah 2021). The high-value targeting
strategy in Pakistan, by this time, had successfully identified and
killed the most threatening terrorists. Only lower-tiered targets would
have remained and their level of threat -- an evaluation of their intent
and capability to harm -- would have likely diminished as well. This
could have encouraged fewer strikes due to the lack of acceptable
targets rather than greater restraint and scrutiny in targeting. The
evidence suggests otherwise. We do not observe a discontinuity in strike
frequency at these cutoffs nor do we see evidence that non-civilian
(combatant) casualties were impacted by exposure to the near certainty
standard.

\hypertarget{economic-implications}{%
\subsection{Economic Implications}\label{economic-implications}}

Our primary analysis estimates an average treatment effect across the
bandwidth of our RD specification. While useful in providing a causal
estimate of the policy's impact in reducing civilian casualties in
Pakistan, the RD analysis does not provide an estimate for the long-run
gains resulting from adoption of the near certainty standard. We follow
a primary assumption of our specification -- as-if randomness of
civilian casualties from strikes -- to project the potential for
civilian deaths during U.S. drone strikes that occurred after the Obama
administration's implementation of a tighter targeting protocol.

We leverage as-if randomness to sample and match -- with replacement --
civilian casualty values during strikes authorized by the Obama
administration before the PPG to strikes ``treated'' by the policy. We
run this exercise as a Monte Carlo simulation with 5,000 iterations and
average across matched civilian casualty values for each treated strike.
We find that these values average to 2.8, which aligns with the
pre-policy average for civilian casualties resulting from U.S. drone
strikes authorized under the more permissive reasonable certainty
standard. This outcome aligns with our RD results as (1) the mean
civilian casualties for strikes carried out under near certainty is
statistically non-different from zero and (2) the RD specification
estimates a per strike reduction of 2.6 civilian casualties attributable
to the discontinuity at the cutoff \(c\). We use the results of our
Monte Carlo simulation to calculate the total number of averted civilian
casualties. First, we take the difference between the matched average
civilian casualty values and the true outcomes. Second, we sum across
these strike-level differences. We estimate 320 averted civilian
casualties in Pakistan were attributable to the Obama administration's
adoption of the near certainty standard.

\FloatBarrier

\begin{figure}[ht]

\includegraphics[width=\textwidth,height=3.64583in]{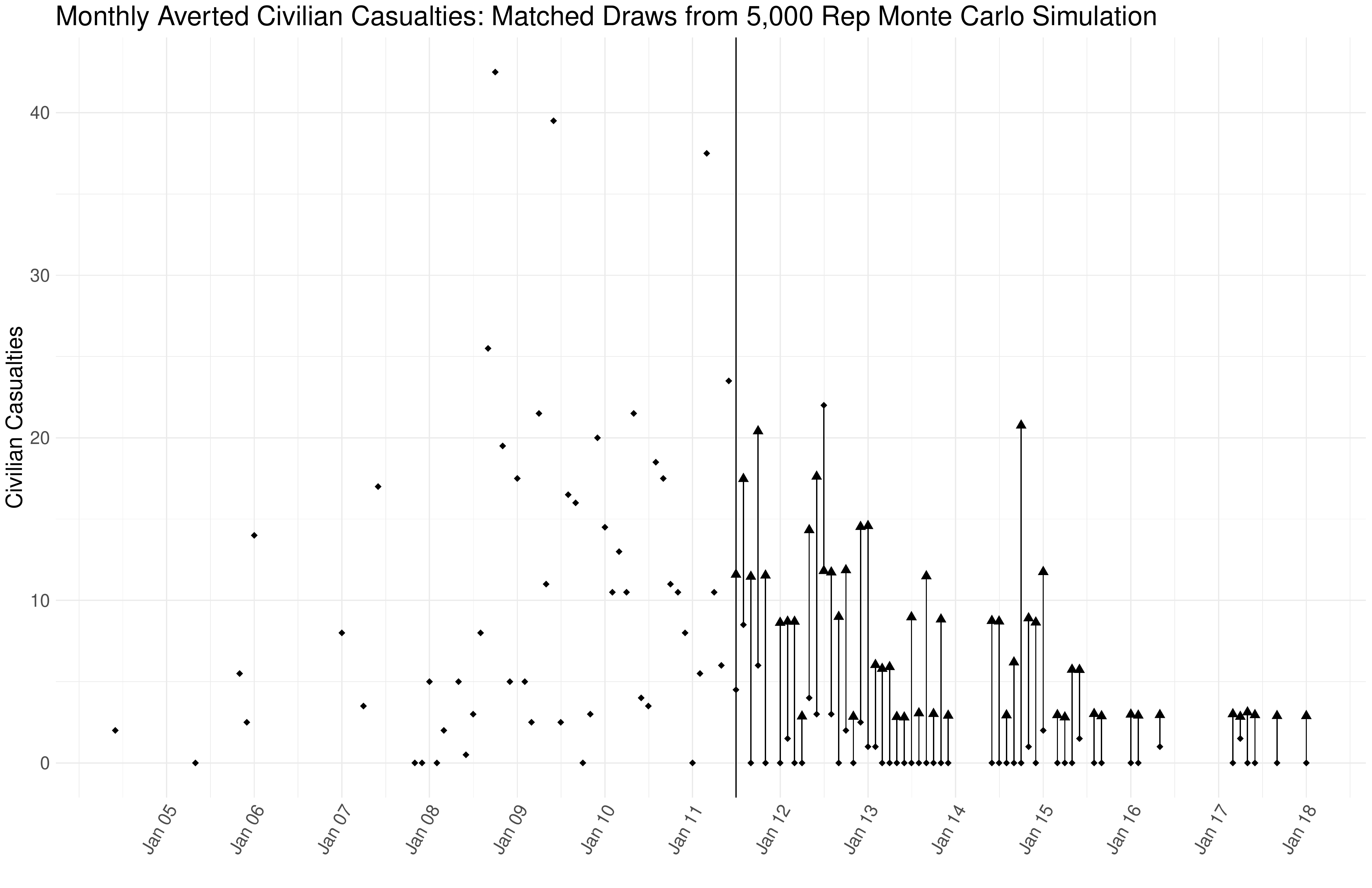}

\caption{Projected civiilian deaths from Monte Carlo simulation displayed as triangular points on plot. Civilians death projections are averaged random draws from 5,000 iteration Monte Carlo simulation at the strike-level, aggregated to monthly totals for visualization. Real civilian casualty values from data are displayed as diamonds in this plot.}

\end{figure}

\FloatBarrier

These averted, potential civilian casualties represent a large financial
and psychological toll had they occurred. Admittedly, this toll is
difficult to quantify and doing so may be morally dubious for some.
Because a preponderance of the drone warfare literature discounts the
economic costs of strikes, we nevertheless attempt to contextualize the
averted loss by providing a VSL calculation based on existing estimates
for individuals in Pakistan. These values represent the ``local tradeoff
rate between money and fatality risk'' (Kniesner and Viscusi 2019). As
such, they offer a reasonable estimate that economists often use to
quantify what this tradeoff is worth given potential for future earnings
and local characteristics. Estimates for a VSL in developing countries
often underestimate this tradeoff rate and therefore vary significantly.
In contrast, a common estimate of a VSL in the United States is
approximately 10 million USD (Viscusi and Masterman 2017). We use low --
200,000 USD -- and high -- 800,000 USD -- estimates for a VSL in
Pakistan to scale our averted death calculation with these VSL values
(Rafiq 2011; Viscusi and Masterman 2017). We estimate a VSL gain between
80 to 260 million USD for the 320 averted civilian casualties that we
attribute to a change in the certainty standard. The body of our
findings carry important policy implications that we discuss below.

\hypertarget{policy-implications}{%
\subsection{Policy Implications}\label{policy-implications}}

Our findings suggest several policy implications relevant to ongoing
debates about whether and how U.S. officials should conduct
counterterrorism drone strikes going forward. First, a heightened level
of political scrutiny allows the United States to enjoy the intended
benefits of drone warfare, such as protections for soldiers and
civilians while surgically removing terrorists. Other countries that
similarly use armed drones, such as France in Mali, have shown that
enhanced oversight is the keystone of capitalizing on the dividends of
UAVs while managing the public controversy of doing so (Vilmer 2017).
U.S. officials openly recognize this benefit of added oversight. Lloyd
Austin, a former four-star U.S. Army General and currently the Secretary
of Defense, recently conceded the U.S. ``must work harder'' to reduce
civilian casualties (Myers 2021). This admission corroborates the
earlier finding by Tirman (2011) that ``despite policies that claim to
protect civilians in war zones,'' U.S. officials ``do not in practice do
so adequately or consistently.''

Second, our findings not only indicate that a near certainty standard
can help prevent civilian casualties during U.S. drone strikes but also
that these shortfalls in policy can impose broad economic and human
costs that are rarely captured by U.S. payments to families of killed
civilians. Policy-makers should consider these when managing the
consequences of tragic errors. It is unclear how the U.S. calibrates its
compensation to families of unintended -- civilian -- victims of
strikes. The available evidence suggests, however, that such ``solatia''
payments are wholly inadequate, even in developing countries such as
Pakistan where the earning potential of citizens is comparatively less
than it is for developed states. A recent report shows Congress has only
authorized 3 million USD a year for compensation payments. In 2019, the
U.S. military made 71 payments to families in Afghanistan and Iraq
ranging from 131 to 35,000 USD (Schmitt 2021a). Even at the low-end of
our VSL calculation for Pakistan -- 200,000 USD -- these payments
dramatically shortchange victims' families, perhaps contributing to
continued grievances (N. Johnston and Bose 2020). While scholars have
often constructed the legitimacy of U.S. drone warfare in terms of the
target (Dill 2015), the economic cost of preventable civilian deaths
should be considered by U.S. officials attempting to rehabilitate
America's image in the wake of egregious mistakes during drone strikes.
This seems especially important if tragic mistakes result in no
accountability among U.S. political and military officials, as was the
case following the Biden administration's botched strike in Afghanistan
as well as a preponderance of previous incidents (Khan 2021; Schmitt
2021b).

According to our analysis, adopting the near certainty standard during
drone strikes in \(declared\) theaters of war can further prevent
civilian casualties. Admittedly, weighing non-combatant immunity in
favor of military necessity exposes soldiers to additional battlefield
risks by restricting commanders' use of drone strikes for
close-air-support, which is designed to defend U.S. forces in combat.
For many war theorists, this is precisely the added risk that reflects
soldiers' martial virtue. Walzer (2015) argues that commanders must
``risk soldiers before they kill civilians.'' Luban adds ``that
accepting \(some\) extra risk to save civilians belongs to the
vocational code of the soldier -- in old-fashioned but still relevant
language it is part of the soldier's code of honor'' (Benbaji 2014). A
handful of military officials have adopted this logic operationally.
Retired U.S. Army General Stanley McChrystal introduced a valor award
for ``courageous restraint'' while leading counterinsurgency operations
in Afghanistan to help institutionalize soldiers' heightened liability
to be harmed while protecting civilians. Many more defense officials,
however, do not endorse tying commanders' hands to protect U.S. forces.
Indeed, McChrystal's initiative failed shortly following its
implementation because it achieved his intent too well: greater numbers
of civilians were saved but at the price of more U.S. casualties (Felter
and Shapiro 2017). If the past is prologue, then, we suspect the
Pentagon will also object to tighter targeting protocols on drone
strikes during active conflicts, though the gains for civilian
protection are unequivocal given our analysis, which is consistent with
other recent studies (Khan 2021).

Finally, our analysis suggests that the U.S. should renew the debate for
the global governance of armed drones, especially if officials recognize
the likelihood of civilian casualties following strikes without greater
oversight. While regulatory mechanisms do exist to counter the
proliferation of the most advanced UAVs, such as the Missile Technology
Control Regime (MTCR), they are feckless and generally lack teeth in
adjudicating the necessary sanctions on violating countries (Mistry
2003). The U.S. is now the most flagrant violator of the MTCR and allows
U.S.-based manufacturers of armed drones such as General Atomics to
directly negotiate export contracts with foreign buyers with little
recourse beyond public criticisms (Stohl and Dick 2021). The frailty of
existing regulatory mechanisms has caused some experts to advocate for
an entirely new approach to regulation optimized to limit the
proliferation of drone technology rather than the platforms themselves.
This is unlikely to be popular given the lucrative profits at stake
(Lushenko, Bose, and Maley 2022). India, for instance, is reportedly
finalizing a 3 billion USD purchase order for 30 MQ-9B SeaGuardian
drones built by General Atomics (Pandit 2021). Without broader
oversight, the international trade of armed drones will contribute to
countries' continued abuse of drone strikes, imposing additional costs
on innocent civilians caught in the crossfire (Lushenko, Bose, and Maley
2022).

\break

\hypertarget{references}{%
\section{References}\label{references}}

\hypertarget{refs}{}
\begin{CSLReferences}{1}{0}
\leavevmode\vadjust pre{\hypertarget{ref-aaronson_precision_2014}{}}%
Aaronson, Mike, Wali Aslam, Tom Dyson, and Regina Rauxloh, eds. 2014.
\emph{Precision Strike Warfare and International Intervention}. 0th ed.
Routledge. \url{https://doi.org/10.4324/9781315850528}.

\leavevmode\vadjust pre{\hypertarget{ref-abizaid_task_2015}{}}%
Abizaid, Gen John P, and Rosa Brooks. 2015. {``{TASK} {FORCE}
{CO}-{CHAIRS},''} April, 81.

\leavevmode\vadjust pre{\hypertarget{ref-aikins_times_2021}{}}%
Aikins, Matthieu, Christopher Koettl, Evan Hill, and Eric Schmitt. 2021.
{``Times Investigation: In u.s. Drone Strike, Evidence Suggests No
{ISIS} Bomb.''} September 10, 2021.
\url{https://www.nytimes.com/2021/09/10/world/asia/us-air-strike-drone-kabul-afghanistan-isis.html}.

\leavevmode\vadjust pre{\hypertarget{ref-alston_promotion_2010}{}}%
Alston, Philip. 2010. {``Promotion and Protection of All Human Rights,
Civil, Political, Economic, Social and Cultural Rights, Including the
Right to Development.''} Koninklijke Brill {NV}.
\url{https://doi.org/10.1163/2210-7975_HRD-9970-2016149}.

\leavevmode\vadjust pre{\hypertarget{ref-anderson_subways_2014}{}}%
Anderson, Michael L. 2014. {``Subways, Strikes, and Slowdowns: The
Impacts of Public Transit on Traffic Congestion.''} \emph{American
Economic Review} 104 (9): 2763--96.
\url{https://doi.org/10.1257/aer.104.9.2763}.

\leavevmode\vadjust pre{\hypertarget{ref-aslam_united_2013}{}}%
Aslam, Wali. 2013. \emph{The United States and Great Power
Responsibility in International Society: Drones, Rendition and
Invasion}. New International Relations. Milton Park, Abingdon, Oxon ;
New York: Routledge.

\leavevmode\vadjust pre{\hypertarget{ref-auffhammer_clearing_2011}{}}%
Auffhammer, Maximilian, and Ryan Kellogg. 2011. {``Clearing the Air? The
Effects of Gasoline Content Regulation on Air Quality.''} \emph{American
Economic Review} 101 (6): 2687--2722.
\url{https://doi.org/10.1257/aer.101.6.2687}.

\leavevmode\vadjust pre{\hypertarget{ref-b_derosa_accountability_2021}{}}%
B. DeRosa, Mary, and Mitt Regan. 2021. {``Accountability for Targeted
Killing.''} In \emph{Counter-Terrorism}, 61--76. Edward Elgar
Publishing. \url{https://doi.org/10.4337/9781800373075.00012}.

\leavevmode\vadjust pre{\hypertarget{ref-bai_estimating_1998}{}}%
Bai, Jushan, and Pierre Perron. 1998. {``Estimating and Testing Linear
Models with Multiple Structural Changes.''} \emph{Econometrica} 66 (1):
47. \url{https://doi.org/10.2307/2998540}.

\leavevmode\vadjust pre{\hypertarget{ref-banka_killing_2018}{}}%
Banka, Andris, and Adam Quinn. 2018. {``Killing Norms Softly: {US}
Targeted Killing, Quasi-Secrecy and the Assassination Ban.''}
\emph{Security Studies} 27 (4): 665--703.
\url{https://doi.org/10.1080/09636412.2018.1483633}.

\leavevmode\vadjust pre{\hypertarget{ref-benbaji_reading_2014}{}}%
Benbaji, Yitzhak, ed. 2014. \emph{Reading Walzer}. 1 {[}edition{]}.
London ; New York: Routledge, Taylor \& Francis Group.

\leavevmode\vadjust pre{\hypertarget{ref-chen_green_2012}{}}%
Chen, Yihsu, and Alexander Whalley. 2012. {``Green Infrastructure: The
Effects of Urban Rail Transit on Air Quality.''} \emph{American Economic
Journal: Economic Policy} 4 (1): 58--97.
\url{https://doi.org/10.1257/pol.4.1.58}.

\leavevmode\vadjust pre{\hypertarget{ref-clark_legitimacy_2005}{}}%
Clark, Ian. 2005. \emph{Legitimacy in International Society}. Oxford ;
New York: Oxford University Press.

\leavevmode\vadjust pre{\hypertarget{ref-crawford_accountability_2013}{}}%
Crawford, Neta C. 2013. \emph{Accountability for Killing: Moral
Responsibility for Collateral Damage in America's Post-9/11 Wars}.
Oxford: Oxford University Press.

\leavevmode\vadjust pre{\hypertarget{ref-cronin_why_2020}{}}%
Cronin, Audrey Kurth. 2020. {``Why Drones Fail.''} May 1, 2020.
\url{https://www.foreignaffairs.com/articles/somalia/2013-06-11/why-drones-fail}.

\leavevmode\vadjust pre{\hypertarget{ref-davis_effect_2008}{}}%
Davis, Lucas W. 2008. {``The Effect of Driving Restrictions on Air
Quality in Mexico City.''} \emph{Journal of Political Economy} 116 (1):
38--81. \url{https://doi.org/10.1086/529398}.

\leavevmode\vadjust pre{\hypertarget{ref-dill_legitimate_2015}{}}%
Dill, Janina. 2015. \emph{Legitimate Targets? Social Construction,
International Law and {US} Bombing}. Cambridge Studies in International
Relations 133. Cambridge, United Kingdom: Cambridge University Press.

\leavevmode\vadjust pre{\hypertarget{ref-ditzen_testing_2021}{}}%
Ditzen, Jan, Yiannis Karavias, and Joakim Westerlund. 2021. {``Testing
and Estimating Structural Breaks in Time Series and Panel Data in
Stata.''} October 28, 2021. \url{http://arxiv.org/abs/2110.14550}.

\leavevmode\vadjust pre{\hypertarget{ref-fair_elite_2016}{}}%
Fair, Christine, and Ali Hamza. 2016. {``From Elite Consumption to
Popular Opinion: Framing of the {US} Drone Program in Pakistani
Newspapers.''} \emph{Small Wars \& Insurgencies} 27 (4): 578--607.
\url{https://doi.org/10.1080/09592318.2016.1189491}.

\leavevmode\vadjust pre{\hypertarget{ref-felter_limiting_2017}{}}%
Felter, Joseph H., and Jacob N. Shapiro. 2017. {``Limiting Civilian
Casualties as Part of a Winning Strategy: The Case of Courageous
Restraint.''} \emph{Daedalus} 146 (1): 44--58.
\url{https://doi.org/10.1162/DAED_a_00421}.

\leavevmode\vadjust pre{\hypertarget{ref-friedersdorf_obama_2016}{}}%
Friedersdorf, Conor. 2016. {``The Obama Administration's Drone-Strike
Dissembling.''} March 14, 2016.
\url{https://www.theatlantic.com/politics/archive/2016/03/the-obama-administrations-drone-strike-dissembling/473541/}.

\leavevmode\vadjust pre{\hypertarget{ref-hausman_regression_2018}{}}%
Hausman, Catherine, and David S Rapson. 2018. {``Regression
Discontinuity in Time: Considerations for Empirical Applications,''} 23.

\leavevmode\vadjust pre{\hypertarget{ref-imbens_regression_2008}{}}%
Imbens, Guido W., and Thomas Lemieux. 2008. {``Regression Discontinuity
Designs: A Guide to Practice.''} \emph{Journal of Econometrics} 142 (2):
615--35. \url{https://doi.org/10.1016/j.jeconom.2007.05.001}.

\leavevmode\vadjust pre{\hypertarget{ref-jaffer_drone_2016}{}}%
Jaffer, Jameel, ed. 2016. \emph{The Drone Memos: Targeted Killing,
Secrecy, and the Law}. New York: The New Press.

\leavevmode\vadjust pre{\hypertarget{ref-johnston_violence_2020}{}}%
Johnston, Nicolas, and Srinjoy Bose. 2020. {``Violence, Power and
Meaning: The Moral Logic of Terrorism.''} \emph{Global Policy} 11 (3):
315--25. \url{https://doi.org/10.1111/1758-5899.12784}.

\leavevmode\vadjust pre{\hypertarget{ref-johnston_impact_2016}{}}%
Johnston, Patrick, and Anoop Sarbahi. 2016. {``The Impact of {US} Drone
Strikes on Terrorism in Pakistan.''} \emph{International Studies
Quarterly} 60 (2): 203--19. \url{https://doi.org/10.1093/isq/sqv004}.

\leavevmode\vadjust pre{\hypertarget{ref-khan_hidden_2021}{}}%
Khan, Azmat. 2021. {``Hidden Pentagon Records Reveal Patterns of Failure
in Deadly Airstrikes.''} \emph{The New York Times}, December.
\url{https://www.nytimes.com/interactive/2021/12/18/us/airstrikes-pentagon-records-civilian-deaths.html}.

\leavevmode\vadjust pre{\hypertarget{ref-king_designing_1994}{}}%
King, Gary, Robert O. Keohane, and Sidney Verba. 1994. \emph{Designing
Social Inquiry: Scientific Inference in Qualitative Research}.
Princeton, N.J: Princeton University Press.

\leavevmode\vadjust pre{\hypertarget{ref-kniesner_value_2019}{}}%
Kniesner, Thomas J., and W. Kip Viscusi. 2019. {``The Value of a
Statistical Life.''} In \emph{Oxford Research Encyclopedia of Economics
and Finance}, by Thomas J. Kniesner and W. Kip Viscusi. Oxford
University Press.
\url{https://doi.org/10.1093/acrefore/9780190625979.013.138}.

\leavevmode\vadjust pre{\hypertarget{ref-kreps_drones_2016}{}}%
Kreps, Sarah E. 2016. \emph{Drones: What Everyone Needs to Know}. First
edition. New York, {NY}: Oxford University Press.

\leavevmode\vadjust pre{\hypertarget{ref-kreps_analysis_2021}{}}%
Kreps, Sarah E., and Paul Lushenko. 2021. {``Analysis {\textbar} What
Happens Now to u.s. Counterterrorism Efforts in Afghanistan? Washington
Post.''} September 21, 2021.
\url{https://www.washingtonpost.com/politics/2021/09/21/what-happens-now-us-counterterrorism-efforts-afghanistan/}.

\leavevmode\vadjust pre{\hypertarget{ref-liddick_no_2021}{}}%
Liddick, E. M. 2021. {``No Legal Objection, Per Se.''} April 21, 2021.
\url{http://warontherocks.com/2021/04/no-legal-objection-per-se/}.

\leavevmode\vadjust pre{\hypertarget{ref-lindsay_information_2020}{}}%
Lindsay, Jon R. 2020. \emph{Information Technology and Military Power}.
Cornell Studies in Security Affairs. Ithaca ; London: Cornell University
Press.

\leavevmode\vadjust pre{\hypertarget{ref-lushenko_75th_2018}{}}%
Lushenko, Paul. 2018. {``The 75th Ranger Regiment Military Intelligence
Battalion.''} \emph{{MILITARY} {REVIEW}}, 12.

\leavevmode\vadjust pre{\hypertarget{ref-lushenko_drones_2022}{}}%
Lushenko, Paul, Srinjoy Bose, and William Maley, eds. 2022. \emph{Drones
and Global Order: Implications of Remote Warfare for International
Society}. Contemporary Security Studies. Abingdon, Oxon ; New York,
{NY}: Routledge.

\leavevmode\vadjust pre{\hypertarget{ref-martin_drone_2017}{}}%
Martin, Geoff, and Erin Steuter. 2017. \emph{Drone Nation: The Political
Economy of America's New Way of War}. Lanham, Maryland: Lexington Books.

\leavevmode\vadjust pre{\hypertarget{ref-martin_predator_2010}{}}%
Martin, Matt, and Charles Sasser. 2010. \emph{Predator: The
Remote-Control Air War over Iraq and Afghanistan: A Pilot's Story}.
Minneapolis, {MN}: Zenith Press.

\leavevmode\vadjust pre{\hypertarget{ref-mistry_containing_2003}{}}%
Mistry, Dinshaw. 2003. \emph{Containing Missile Proliferation: Strategic
Technology, Security Regimes, and International Cooperation in Arms
Control}. Seattle: University of Washington Press.

\leavevmode\vadjust pre{\hypertarget{ref-moyn_humane_2021}{}}%
Moyn, Samuel. 2021. \emph{Humane: How the United States Abandoned Peace
and Reinvented War}.
\url{https://www.overdrive.com/search?q=7AD56903-BD24-40A3-8EA3-7AD86E8D59C8}.

\leavevmode\vadjust pre{\hypertarget{ref-myers_we_2021}{}}%
Myers, Meghann. 2021. {``{`We Must Work Harder,'} {SECDEF} Says as
Pentagon Grapples with Civilian Casualties of Airstrikes. Military
Times.''} November 17, 2021.
\url{https://www.militarytimes.com/news/pentagon-congress/2021/11/17/we-must-work-harder-secdef-says-as-pentagon-grapples-with-civilian-casualties-of-airstrikes/}.

\leavevmode\vadjust pre{\hypertarget{ref-odom_intelligence_2008}{}}%
Odom, William E. 2008. {``Intelligence Analysis.''} \emph{Intelligence
and National Security} 23 (3): 316--32.
\url{https://doi.org/10.1080/02684520802121216}.

\leavevmode\vadjust pre{\hypertarget{ref-owens_us_2011}{}}%
Owens, Mackubin Thomas. 2011. \emph{{US} Civil-Military Relations After
9/11: Renegotiating the Civil-Military Bargain}. New York: Continuum.

\leavevmode\vadjust pre{\hypertarget{ref-page_drones_2021}{}}%
Page, James Michael, and John Williams. 2021. {``Drones, Afghanistan,
and Beyond: Towards Analysis and Assessment in Context.''}
\emph{European Journal of International Security}, October, 1--21.
\url{https://doi.org/10.1017/eis.2021.19}.

\leavevmode\vadjust pre{\hypertarget{ref-pandit_3_2021}{}}%
Pandit, Rajat. 2021. {``\$3 Billion Predator Drone Deal: India Seeks
Clarity from {US} on Price, Tech Transfer {\textbar} India News - Times
of India.''} August 25, 2021.
\url{https://timesofindia.indiatimes.com/india/3bn-predator-drone-deal-india-seeks-clarity-from-us-on-price-tech-transfer/articleshow/85611977.cms}.

\leavevmode\vadjust pre{\hypertarget{ref-plaw_practice_2011}{}}%
Plaw, Avery, Matthew S Fricker, and Brian Glyn Williams. 2011.
{``Practice Makes Perfect?''} 5: 20.

\leavevmode\vadjust pre{\hypertarget{ref-primoratz_civilian_2007}{}}%
Primoratz, Igor, ed. 2007. \emph{Civilian Immunity in War}. Oxford ; New
York: Oxford University Press.

\leavevmode\vadjust pre{\hypertarget{ref-rafiq_estimating_2011}{}}%
Rafiq, Muhammad. 2011. \emph{Estimating the Value of Statistical Life in
Pakistan}. {SANDEE} Working Paper 63. Kathmandu: South Asian Network for
Development; Environmental Economics.

\leavevmode\vadjust pre{\hypertarget{ref-noauthor_remarks_2013}{}}%
{``Remarks by the President at the National Defense University.''} 2013.
May 23, 2013.
\url{https://obamawhitehouse.archives.gov/the-press-office/2013/05/23/remarks-president-national-defense-university}.

\leavevmode\vadjust pre{\hypertarget{ref-renic_asymmetric_2020}{}}%
Renic, Neil C. 2020. \emph{Asymmetric Killing: Risk Avoidance, Just War,
and the Warrior Ethos}. New York, {NY}: Oxford University Press.

\leavevmode\vadjust pre{\hypertarget{ref-rigterink_wane_2021}{}}%
Rigterink, Anouk S. 2021. {``The Wane of Command: Evidence on Drone
Strikes and Control Within Terrorist Organizations.''} \emph{American
Political Science Review} 115 (1): 31--50.
\url{https://doi.org/10.1017/S0003055420000908}.

\leavevmode\vadjust pre{\hypertarget{ref-riza_killing_2013}{}}%
Riza, M. Shane. 2013. \emph{Killing Without Heart: Limits on Robotic
Warfare in an Age of Persistent Conflict}. First edition. Washington,
D.C: Potomac Books.

\leavevmode\vadjust pre{\hypertarget{ref-sanger_4_2017}{}}%
Sanger, David E. 2017. {``4 Cyber, Drones, and Secrecy,''} 19.

\leavevmode\vadjust pre{\hypertarget{ref-savage_us_2016}{}}%
Savage, Charlie. 2016. {``U.s. Releases Rules for Airstrike Killings of
Terror Suspects - the New York Times.''} August 6, 2016.
\url{https://www.nytimes.com/2016/08/07/us/politics/us-releases-rules-for-airstrike-killings-of-terror-suspects.html}.

\leavevmode\vadjust pre{\hypertarget{ref-schmitt_us_2021}{}}%
Schmitt, Eric. 2021a. {``U.s. Pledges to Pay Family of Those Killed in
Botched Kabul Drone Strike.''} \emph{The New York Times}, October.
\url{https://www.nytimes.com/2021/10/15/us/politics/kabul-drone-strike-victims-payment.html}.

\leavevmode\vadjust pre{\hypertarget{ref-schmitt_no_2021}{}}%
---------. 2021b. {``No u.s. Troops Will Be Punished for Deadly Kabul
Strike, Pentagon Chief Decides.''} December 13, 2021.
\url{https://www.nytimes.com/2021/12/13/us/politics/afghanistan-drone-strike.html}.

\leavevmode\vadjust pre{\hypertarget{ref-shah_us_2018}{}}%
Shah, Aqil. 2018. {``Do u.s. Drone Strikes Cause Blowback? Evidence from
Pakistan and Beyond.''} \emph{International Security} 42 (4): 47--84.
\url{https://doi.org/10.1162/isec_a_00312}.

\leavevmode\vadjust pre{\hypertarget{ref-sheehan_routledge_2021}{}}%
Sheehan, Michael A., Erich Marquardt, and Liam Collins. 2021.
\emph{Routledge Handbook of u.s. Counterterrorism and Irregular Warfare
Operations}. 1st ed. London: Routledge.
\url{https://doi.org/10.4324/9781003164500}.

\leavevmode\vadjust pre{\hypertarget{ref-stohl_new_2021}{}}%
Stohl, Rachel, and Shannon Dick. 2021. {``A New Agenda for {US} Drone
Policy and the Use of Lethal Force • Stimson Center. Stimson Center.''}
April 29, 2021.
\url{https://www.stimson.org/2021/a-new-agenda-for-us-drone-policy-and-the-use-of-lethal-force/}.

\leavevmode\vadjust pre{\hypertarget{ref-swan_drone_2019}{}}%
Swan, Ryan. 2019. {``Drone Strikes: An Overview, Articulation and
Assessment of the United States' Position Under International Law.''}
{LLNL}-{TR}--800242, 1580679, 976830.
\url{https://doi.org/10.2172/1580679}.

\leavevmode\vadjust pre{\hypertarget{ref-tirman_deaths_2011}{}}%
Tirman, John. 2011. \emph{The Deaths of Others: The Fate of Civilians in
America's Wars}. New York: Oxford University Press.

\leavevmode\vadjust pre{\hypertarget{ref-turse_how_2021}{}}%
Turse, Nick. 2021. {``How Biden Is Trying to Rebrand the Drone War.''}
October 25, 2021.
\url{https://responsiblestatecraft.org/2021/10/25/how-biden-is-trying-to-rebrand-the-drone-war/}.

\leavevmode\vadjust pre{\hypertarget{ref-ullah_terrorism_2021}{}}%
Ullah, Imdad. 2021. \emph{Terrorism and the {US} Drone Attacks in
Pakistan: Killing First}. 1st ed. Routledge.
\url{https://doi.org/10.4324/9781003145486}.

\leavevmode\vadjust pre{\hypertarget{ref-vilmer_french_2017}{}}%
Vilmer, Jean-Baptiste Jeangene. 2017. {``The French Turn to Armed
Drones. War on the Rocks.''} September 22, 2017.
\url{https://warontherocks.com/2017/09/the-french-turn-to-armed-drones/}.

\leavevmode\vadjust pre{\hypertarget{ref-viscusi_income_2017}{}}%
Viscusi, W. Kip, and Clayton J. Masterman. 2017. {``Income Elasticities
and Global Values of a Statistical Life.''} \emph{Journal of
Benefit-Cost Analysis} 8 (2): 226--50.
\url{https://doi.org/10.1017/bca.2017.12}.

\leavevmode\vadjust pre{\hypertarget{ref-walzer_just_2015}{}}%
Walzer, Michael. 2015. \emph{Just and Unjust Wars: A Moral Argument with
Historical Illustrations}. Fifth edition. New York: Basic Books, a
member of the Perseus Books Group.

\leavevmode\vadjust pre{\hypertarget{ref-zenko_obamas_2017}{}}%
Zenko, Micah. 2017. {``Obama's Final Drone Strike Data.''} January 20,
2017. \url{https://www.cfr.org/blog/obamas-final-drone-strike-data}.

\end{CSLReferences}

\break

\appendix

\hypertarget{appendix}{%
\section{Appendix}\label{appendix}}

\hypertarget{errata}{%
\subsection{Errata}\label{errata}}

\hypertarget{manuscript-meta}{%
\subsubsection{Manuscript Meta}\label{manuscript-meta}}

8385 words, 7 in-text figures and 2 in-text tables.

All tables are attached as {[}TeX{]} files and all figures are attached
as {[}PDF{]} files.

The references for this manuscript were generated from
{[}ds\_updated.bib{]}, attached.

\hypertarget{acknowledgments}{%
\subsubsection{Acknowledgments}\label{acknowledgments}}

The authors thank Jacob Shapiro, Patrick Carlin, Nicolas Ziebarth,
Nicolas Bottan, Nicholas Sanders, John Cawley, Barton Willage, Lt. Col.
(U.S. Army) Keith Carter, Grace Phillips, Lt. Col. (U.S. Army) Anthony
Williams, Tom Pepinsky, Alexandra Blackman, and participants of the
Cornell PIERS seminar for their helpful comments and guidance. Richard
Li provided excellent research assistance.

\hypertarget{funding-statement}{%
\subsubsection{Funding Statement}\label{funding-statement}}

The authors did not receive funding for this research.

\hypertarget{orcid}{%
\subsubsection{ORCID}\label{orcid}}

The authors' ORCID links are provided below:

Raman: \href{orcid.org/0000-0002-5984-5417}{0000-0002-5984-5417}

Lushenko: \href{orcid.org/0000-0002-9299-9709}{0000-0002-9299-9709}

Kreps: \href{orcid.org/0000-0002-0924-4234}{0000-0002-0924-4234}

\break

\hypertarget{supplemental-information}{%
\subsection{Supplemental Information}\label{supplemental-information}}

\hypertarget{elite-interviews}{%
\subsubsection{Elite Interviews}\label{elite-interviews}}

In defining the sharp cutoff for our RDiT specification, we conducted
several interviews with Obama-era officials, including senior
policy-makers and intelligence analysts responsible for the
implementation and oversight of the near-certainty standard. On November
5, 2021, we interviewed Luke Hartig. As Obama's Senior Director for
Counterterrorism at the National Security Council, Hartig was a
principal author and manager of the near certainty standard risk
mitigation process. We conducted a follow-up interview with Hartig on
December 6, 2021. Additionally, we interviewed Jamil Jaffer on December
10, 2021. Jaffer served as the Senior Counsel to the House Permanent
Select Committee on Intelligence during the Obama administration. He
also served on the leadership team of the Senate Foreign Relations
Committee as Chief Counsel and Senior Advisor under Chairman Bob Corker
(R-TN), where he worked on key national security and foreign policy
issues, including leading the drafting of the proposed Authorization for
the Use of Military Force (AUMF) against the Islamic State in 2014 and
2015, the AUMF against Syria in 2013, and revisions to the 9/11 AUMF
against al-Qaeda.

\hypertarget{methods}{%
\subsubsection{Methods}\label{methods}}

We model the impacts of the Obama administration's policy in several
specifications to evaluate the relationship between the change in the
certainty standard and civilian casualties. We initially perform a
structural break analysis to estimate any points of structural
discontinuity in our time series data. We then perform a one-way ANOVA
to evaluate the relationship between exposure to the policy on our
civilian casualty outcomes, using the estimated structural breakpoint.
Drawing on the previous two analyses, we fit a regression discontinuity
model to provide a causal estimate for the impact of the near certainty
standard on our outcomes in Pakistan during the Obama administration. We
offer a set of robustness checks that validate the key assumptions of
our RDiT design. Finally, we perform a VSL calculation on averted
civilian casualties attributable to the Obama administration's near
certainty standard.

\hypertarget{stuctural-break-and-anova-results}{%
\subsubsection{Stuctural Break and ANOVA
Results}\label{stuctural-break-and-anova-results}}

We begin our analysis by investigating two questions. First, we are
interested if there exists a structural break in the trend of civilian
casualties in our time series. Though Obama announced the near certainty
standard on May 23, 2013, the implementation took place over the
previous two years leading up to the announcement. We follow Ditzen,
Karavias, and Westerlund (2021) and estimate structural breakpoints in
our data and then test the significance of those estimates following Bai
and Perron (1998). We estimate a calendar time structural break in the
trend of civilian casualties within our implementation window at July
{[}1{]}, 2011 {[}p = .01, 95\% CI: May-Aug 2011{]}.

Secondly, we are interested in if -- conditional on exposure to the near
certainty standard that accompanied this policy -- there is a
significant difference in the means of treated and non-treated
observations. We fit a series of one-way ANOVA models on the following
outcomes at the strike-level: (1) civilian casualties, (2) strike
precision, and (3) civilian casualties per strike. We find a
statistically significant difference in the means of these outcomes
conditional on exposure to the near certainty standard at both our
structural break analysis estimate (July {[}1{]}, 2011) and the Obama
administration's public announcement of the policy (May 23, 2013). We
use differences across policy exposure to inform the potential outcomes
framework for our RDiT specification and as such, model each of these
ANOVA outcomes in our main RDiT specification.

\bibliographystyle{unsrt}
\bibliography{ds\_updated.bib}

\begin{thebibliography}{100}
\bibitem{noauthor_us_nodate}
A u.s. drone strike in kabul was a ‘tragic mistake.’ - the washington post.
\end{thebibliography}

\end{document}